\documentclass[usenatbib]{mn2e}
\usepackage{graphicx,times}
\usepackage{amsmath,amssymb}

\title[Role of Gas in Spiral Galaxies]{The Role of Gas In Maintaining Quasi-Steady Spiral Structure in Stellar Disks}
\author[Sukanya Chakrabarti]{Sukanya Chakrabarti$^{1,2}$\thanks{Email: schakrab@cfa.harvard.edu}\\
$^1$Harvard-Smithsonian Center for Astrophysics, 60 Garden Street, Cambridge, MA 02138 USA\\
$^2$Institute for Theory \& Computation Postdoctoral Fellow}

\begin{document}
\maketitle

%\label{firstpage}

\begin{abstract}  We study the dynamical evolution of spiral structure in the stellar
disks of isolated galaxies using high resolution Smoothed Particle Hydrodynamics
(SPH) simulations that treat the evolution of gas, stars, and dark matter
self-consistently.  We focus this study on the question of self-excited spiral structure in the
stellar disk and investigate the dynamical coupling between the cold, dissipative gaseous
component and the stellar component.  We find that angular momentum transport from the gas
to the stars inside of corotation leads to a roughly time-steady spiral structure in the
stellar disk.  To make this point clear, we contrast these results with otherwise identical  simulations that do not include a cold gaseous component that is able to cool radiatively and dissipate energy, and find that spiral structure, when it is incipient, dies out more rapidly in simulations that do not include gas.  We also employ a standard star formation prescription to convert gas into
stars and find that our results hold for typical gas consumption time scales that 
are in accord with the Kennicutt-Schmidt relation.  We therefore attribute the long-lived roughly
 time steady spiral structure in the stellar disk to the dynamical coupling between the gas and the stars and the resultant torques that the self-gravitating gaseous disk is able to exert on the stars due to an azimuthal phase shift between the collisionless and dissipative components.  

\end{abstract} 

%\begin{keywords}
%galaxies: formation---galaxies: spiral---stars:formation}
%\end{keywords}


\section{Introduction}

Spiral galaxies present two faces to observers.  When seen in blue light, these galaxies look disordered, with spurs, branches and feather-like structures seen even in grand design galaxies, while K-band images show that these galaxies possess smooth, usually two-armed spiral patterns.  We study here the dynamical interaction between the Population I component seen in blue light and the older Population II component revealed in the longer wavelengths, specifically with the intent of understanding how the interaction between the gas and the stars affects the time evolution of spiral structure in the stellar disk.

The idea that the stellar spiral structure is not a material phenomenon, but a roughly time-steady density wave rotating at a pattern speed was developed in the seminal work of Lin \& Shu (1964; henceforth LS64).  Toomre (1964) gave the minimum radial stellar velocity disperion required to suppress the onset of local instabilities in a razor-thin stellar disk stabilized by rotation and the velocity dispersion of stars on large and small scales respectively.  Much of the early numerical work of the study of spiral structure utilized N-body simulations to study the growth of spiral instabilities in the collionsionless components of galaxies.  Transient spiral structure was evinced in the early N-body simulations of Hohl (1970), as well as in the later works of Sellwood \& Carlberg (1984; henceforth SC84).  Toomre (1981) showed that a shearing disk with a supply of leading waves will swing amplify these leading disturbances to trailing disturbances, and was one of the earliest works to emphasize the role of cosmological accretion of gas.  Tidal disturbances or minor mergers can torque the stellar disk and create transient spiral structure (Hernquist \& Mihos 1995).  Some of these early influential works noted that the dissipative nature of gas may provide a self-regulatory influence in the combined gas-disk system (Lin \& Shu 1964; Bertin \& Lin 1996), with dissipation in the gas layer serving to continually regenerate the shearing spiral instability (Goldreich \& Lynden-Bell 1965). 

Contemporary numerical investigations of spiral structure have focused mainly on the response of a gas disk when driven by a spiral potential, to varying levels of physical detail and numerical sophistication.  Kim \& Ostriker (2002) carried out local magnetohydrodynamic (MHD) simulations to study the development of spiral substructure in a gas disk forced by a spiral potential.  Chakrabarti, Laughlin \& Shu (2003) (henceforth CLS) studied the development of spiral substructure due to ultraharmonic resonances using global simulations which incorporate the self-gravity of the gas and found that inter-arm features of long azimuthal extent occurred at resonance locations, which they called ``branches''.  CLS also found that when the gas disk is forced by a time-steady, but nonlinear spiral potential (which is composed of many Fourier components), overlap of resonances produced a highly time-dependent response in the gas which they speculated may signal the onset of turbulence in the purely hydrodynamic regime.  Shetty \& Ostriker (2006; henceforth SO06) carried out global MHD simulations and found that the magneto-Jeans instability is effective at forming spiral-substructure in low-shear (i.e., near the spiral arms) regions when the magnetic field is large.  The stabilizing effect of magnetic pressure allowed these simulations to be carried out for longer times than those of CLS.  Yanez et al.'s (2007) hydrodynamic simulations also evince the role of ultraharmonic resonances in the development of spiral sub-structure in the gaseous component.  A significant work on the analysis of spiral structure in self-gravitating disks is that of Laughlin et al. (1998) - these authors rigorously show that the primary unstable mode is endemic to the disk and not due to the numerical implementation.

One of the essential points evinced by these simulations of the response of a self-gravitating gas disk to a time-steady stellar potential is that low $Q$ gas disks display a very time-dependent response and continue to do so over many dynamical times (where we refer to the Toomre Q parameter, i.e., $Q=c_{i}\kappa/\pi G \Sigma$, where $c_{i}$ is the sound speed of the relevant component; Toomre 1964).  This is in marked contrast to purely stellar disks which may exhibit transient spiral structure (if they are low $Q$ disks) but the spiral patterns disappear after a few dynamical times.  

Block et al. (1994; 1996) found that many spiral galaxies have a grand design structure in infrared light (Elmegreen \& Elmegreen 1984; Rix \& Zaritsky 1995) while appearing quite flocculent in blue light.  They interpret this discrepancy in morphology to mean that the Population I and Population II components are dynamically decoupled.  Recent multi-wavelength observations have begun to address in detail the long term evolution of spiral structure in nearby galaxies.  Notably, Kendall et al. (2008) infer from analysis of high spatial resolution multi-wavelength photometry of M81 (from the B-band to IRAC bands) a phase shift between the gas and the stars that is suggestive of a long-lived spiral structure in the stellar disk.  Global torques may have a role in fueling star formation over many Gyrs if they are able to transport the large amounts of mass in extended HI disks (Wong \& Blitz 2002) to the inner regions.  These issues require a ${\it global}$ analysis that treats the interplay between gas and stars rather than treating the response of the gas to an imposed stellar potential as has been studied previously (CLS; SO06, among other previous papers).

In this paper, we employ high resolution three-dimensional simulations performed with the Smoothed Particle Hydrodynamics (SPH) code GADGET-2 (Springel 2005) to simulate the evolution of $\it{self-excited~spiral~structure}$ in the stellar disks of galaxies that are evolving in relative isolation.  
  The recent analysis of the chemical and star formation history of the Milky Way by Colavitti et al. (2008) suggests that the Milky Way has not undergone recent (i.e., after $z \sim 1$) significant merger events.  We study here the problem of the generation and evolution of spiral structure in an isolated galaxy with gas, stars, and a dark matter halo, where these components are free to interact dynamically with each other.  Specifically, by including not only the stars with their gravitational forces and velocity dispersions, but also the cold interstellar gas with its self-gravity, pressure, its ability to cool radiatively and dissipate energy through shocks, and form new stars, we focus our study on the dynamical coupling between the gas and the stars and the torques that the gas is able to exert on the stars, an effect that has not been analyzed in previous studies which have focused on the effect of the stars (whether live or modeled through a potential) on the gas (Hernquist \& Mihos 1995; CLS; S006; Kim \& Ostriker 2007).  We also comment on the coupling with the live halo and its role as a sink of angular momentum, as has been noted in previous studies of bar instabilities (Athanassoula 2003; Martinez-Valpuesta et al. 2006; Berentzen et al. 2007).  We find that the dissipative nature of the gas leads to $\it{sharp}$, $\it{asymmetric}$ azimuthal features, with the gas response leading the stellar response, which allows for a phase shift between these components such that the gas is able torque the stars.  It is this angular momentum transport interior to co-rotation that allows for the stellar disk to be unstable to spiral structure for a longer time (many revolutions) compared to simulations that do not include the dissipative gaseous component.

The paper is organized as follows.  In \S 2, we review the simulation methodology, specifically the initial equilibrium structure of the disks and implementation of radiative cooling, viscosity, and star formation prescriptions for turning gas into stars.  In \S 3, we give our results for simulations that treat the collionsionless components only, and in \S 4, we describe results from simulations that include the time evolution of gas, stars, and a dark matter halo.  
We discuss our results in the context of observations in \S 5, and we conclude in \S 6.  The Appendices are devoted to a detailed exploration of simulation parameters and assumptions that can affect the gas-star angular momentum transport, specifically the artificial bulk viscosity and equation of state for the gas.

\section{Simulation Methodology}

We employ the parallel TreeSPH code GADGET-2 (Springel 2005) to perform simulations of isolated disk galaxies.  GADGET-2 uses an N-body method to follow the evolution of the collionsionless components, and SPH to follow the gaseous component.  Energy and entropy are both conserved in regions free of dissipation.  The simulations reported here (unless otherwise noted) have gravitational softening lengths of $25~\rm pc$ for the gas and stars, and $50~\rm pc$ for the halo.  The number of gas, stellar, and halo particles are $10^{5}$, $1.08 \times 10^{6}$ and $10^{6}$ respectively for our fiducial case.  The total angular momentum of these simulations is conserved to 0.1 \% over 3 Gyr for a time integration error tolerance accuracy of 0.025.  In Appendix A, we give some results of a few cases where we vary the artificial bulk viscosity and the equation of state - these simulations are performed at lower resolution and their specifications are noted in the appendix.  

The implementation of viscosity in GADGET-2 is described in detail in Springel et al. (2001) and Springel (2005), so we briefly review the results here.  GADGET-2 uses a bulk artificial viscosity which is the shear-reduced version (Balsara 1995; Steinmetz 1996) of the standard Monaghan and Gingold (1983) artificial viscosity.  Specifically, the form of the artificial viscosity experienced between particles $i$ and $j$ is taken to be given by:
\begin{equation}
\Pi_{\rm ij}=-\frac{\alpha}{2}\frac{\left(c_{i}+c{j}-3w_{i}\right)w_{j}}{\rho_{ij}}\;,
\end{equation} 
where $\rho_{ij}$ is the arithmetic mean of the SPH density estimator (equation 5 in Springel 2005) of particles $i$ and $j$, $c_{i}$ and $c_{j}$ are the sound speeds, and $w_{i}$ and $w_{j}$ are the relative projected velocities. 
Here, we adopt a fiducial value of 0.75 for the bulk artificial viscosity $\alpha$ parameter.  This value has been empirically recommended by Springel et al. (2001) by comparing the results of the shock tube test to other numerical implementations.  In discontinuities that develop in the gaseous flow, the entropy generation and width of the shock is proportional to this parameter.  We vary this parameter in Appendix A and find that angular momentum transport from the gas to the stars is inhibited for very low values of the artificial bulk viscosity parameter.  For low $\alpha$ values, the effect of dissipation in shocks is prohibited such that the gaseous component becomes more akin to the collionsionless component and the stars no longer exhibit a significant time lag or azimuthal phase shift in their response.

Radiative cooling and heating are implemented in GADGET-2 in a manner similar to Katz, Weinberg \& Hernquist (1996) (henceforth KWH96).  The code implements radiative cooling for an optically thin gas of helium and hydrogen in collisional ionization equilibrium with a user-specified (time-dependent) ultraviolent background.  Radiative cooling allows the gas to dissipate energy and form dense, cold clumps, which as we describe below are prone to star formation.  For the densities resolved in these simulations, radiative cooling due to two-body processes is important, namely, collisional excitation of neutral hydrogen ($\rm H^{0}$) and singly ionized helium ($\rm He^{+}$), collisional ionization of $\rm H^{0}, He^{0}$ and $\rm He^{+}$, standard recombination of $\rm H^{+}, He^{+}$, and $\rm He^{++}$, dielectric recombination of $He^{+}$ and free-free emission.  The adopted cooling rates are given in Table 1 of KWH96.  The heating rate from photoionization can be computed once the intensity of the UV background is specified.  Since this heating rate is very low inside disks in the local universe, we do not make use of this heating rate from a UV background.  

The version of GADGET-2 that we employ uses a sub-resolution model for energy injection from supernovae that is proportional to the star formation rate (Springel \& Hernquist 2003).  In this approach, the ISM is taken to be composed of cold clouds from which stars can form, which are embedded in a pressure-confining hot phase.  The effective equation of state for this two-component gas (Eq. 24 in Springel et al. 2001) is taken to be 0.25 for our fiducial cases.  In essence, employing such a equation of state for the star forming gas allows one to transition from the isothermal case to a stiffer than isothermal equation of state, where in the latter case energy injection from supernovae (which is proportional to the star formation rate) allows one to evolve gas-rich disks for many dynamical times.  We vary the equation of state in Appendix B and study its effect on the gas-star angular momentum transport.

We discuss in this paper both simulations where star formation is turned on and those where star formation is turned off.  For simulations where star formation is turned on, we employ a density dependent Kennicutt-Schmitdt algorithm, to convert gas to stars in these simulations.  The star formation rate is given by:

\begin{equation}
\frac{d\rho_{\star}}{dt}=(1-\beta)\frac{\rho}{t_{\star}}  \;,
\end{equation}
where $\rho$ is the total gas density, $\beta \sim 0.1$ for a Salpeter IMF, and the quantity $t_{\star}$ is given by:

\begin{equation}
t_{\star}(\rho)=t_{0}^{\star}\left(\frac{\rho}{\rho_{\rm th}}\right)^{-1/2} \;,
\end{equation}
where $\rho_{\rm th}$ is fixed by the multiphase model and is the threshold density above which the gas is assumed to be thermally unstable to the onset of a two-phase medium composed of cold star forming clouds embedded in a hot, pressure-confining phase (SH03).   
We take $t_{0}^{\star}$, the gas consumption timescale, to be 4.5 Gyr and 8 Gyr which are intermediate values between that adopted by SH03 and Springel et al. (2005) and agree reasonably well (to within a factor of 2-3) with gas consumption time scales cited by Kennicutt (1998).  We aim to highlight in this paper general trends in the dynamics that are seen across simulations.  Due to the large dynamic range in global galaxy simulations, star formation is currently treated in a prescriptive way.  To minimize dependence of our results on our choice of a star formation prescription, we also include results from simulations where the star formation is turned off.  For the simulation where star formation is turned off, there is no energy injection from supernovae (as SH03's model takes the energy injection rate to be proportional to the star formation rate), and the gas remains isothermal at $10^{4}~\rm K$.  We discuss in a future paper results from various star formation prescriptions (where star formation is mediated by other factors such as shocks, magnetic fields and turbulence) for turning gas into stars.  
\begin{table*}
 \centering
 \begin{minipage}{140mm}
  \caption{Simulation Parameters}
  \begin{tabular}{@{}llrrrrl@{}}
  \hline
   Sim & $f_{\rm gas}$ & $V_{\rm vir}(\rm km/s)$   &$N_{\rm halo}, N_{\rm stars}, N_{\rm gas} (10^{5})$  & EQS & $\nu_{\rm bulk}$ & $t_{0}^{\star} (\rm Gyr)$ \\
       \hline

SpNGNB  &  0   & 160  & $10,10.88,0$    & 0.25  & 0.75  & 0  \\
SpGNSFR & 0.1  & 160  & $10,10.88,1$    & 0.25  & 0.75  & 0  \\
SpGNSFR0& 0.1  & 160  & $10,10.88,1$    & 0.25  & 0     & 0  \\
SpGNSFR1& 0.1  & 160  & $10,10.88,1$    & 0.25  & 1     & 0  \\
SpGKS   & 0.1  & 160  & $10,10.88,1$    & 0.25  & 0.75  & 4.5 \\
SpGKS8  & 0.1  & 160  & $10,10.88,1$    & 0.25  & 0.75  & 8  \\
EQS0LR  & 0.1  & 160  & $1.2,3.6,0.4$   & 0     & 0.75  & 4.5 \\
EQSSTLR & 0.1  & 160  & $1.2,3.6,0.4$   & 0.25  & 0.75  & 4.5 \\
EQS1LR  & 0.1  & 160  & $1.2,3.6,0.4$   & 1     & 0.75  & 4.5 \\

\hline
\end{tabular}
\end{minipage}
\end{table*}

The initial conditions that we choose for the disks represent Milky Way type galaxies and are similar to those described in Cox et al. (2006) and Springel et al. (2005).   Namely, we generate a stable isolated disk galaxy that is initially in equilibrium.   The extended dark matter halo is taken to have a Hernquist (1990) profile, with a concentration parameter of 10.  We model the disk components of gas and stars as exponential disks, with the fiducial case having a scale length of 3 kpc.  The primary simulation that we focus on here has a virial velocity of $160~\rm km/s$, with a rotation curve and mass similar to the Milky Way ($M_{\rm total}=1.3 \times 10^{12} M_{\odot}$ and a stellar mass of $5.4 \times 10^{10} M_{\odot}$ initially).  We do not include a bulge in the simulations reported here (there is no qualitative difference as far as our main results are concerned), and take the baryonic gas mass fraction to be 10 \% for the fiducial case, with the gas distributed initially in an axisymmetric exponential disk having the same scale length as the stars.  (In retrospect, inclusion of an extended gas disk as seen in observations by Wong \& Blitz (2002) would have been more empirically suited)  Values of these parameters for the other simulations are reported in Table 1.  These values yield an initial Toomre Q parameter for the stars that is slightly in excess of unity in the inner regions, and of order unity for the gas.
We have performed resolution studies to confirm that simulations with the given number of particles and softening lengths yield converged results.  The primary simulations that we analyze here are the collisionless SpNGNB simulation, its gaseous analogue (SpGNSFR), and its gaseous, star-forming analogues (SpGKS and SpGKS8). For a given cooling curve and star formation prescription, there are two aspects of the gas physics that can affect the gas-star dynamical coupling in these simulations, namely the artificial bulk viscosity and the equation of state for the gas.  These aspects of the simulation methodology and their effect on angular momentum transport and treated in Appendix A and B respectively.

\section{Simulations of the Collisionless Components of Galaxies}

\begin{figure}
\begin{center}
\includegraphics[scale=0.25]{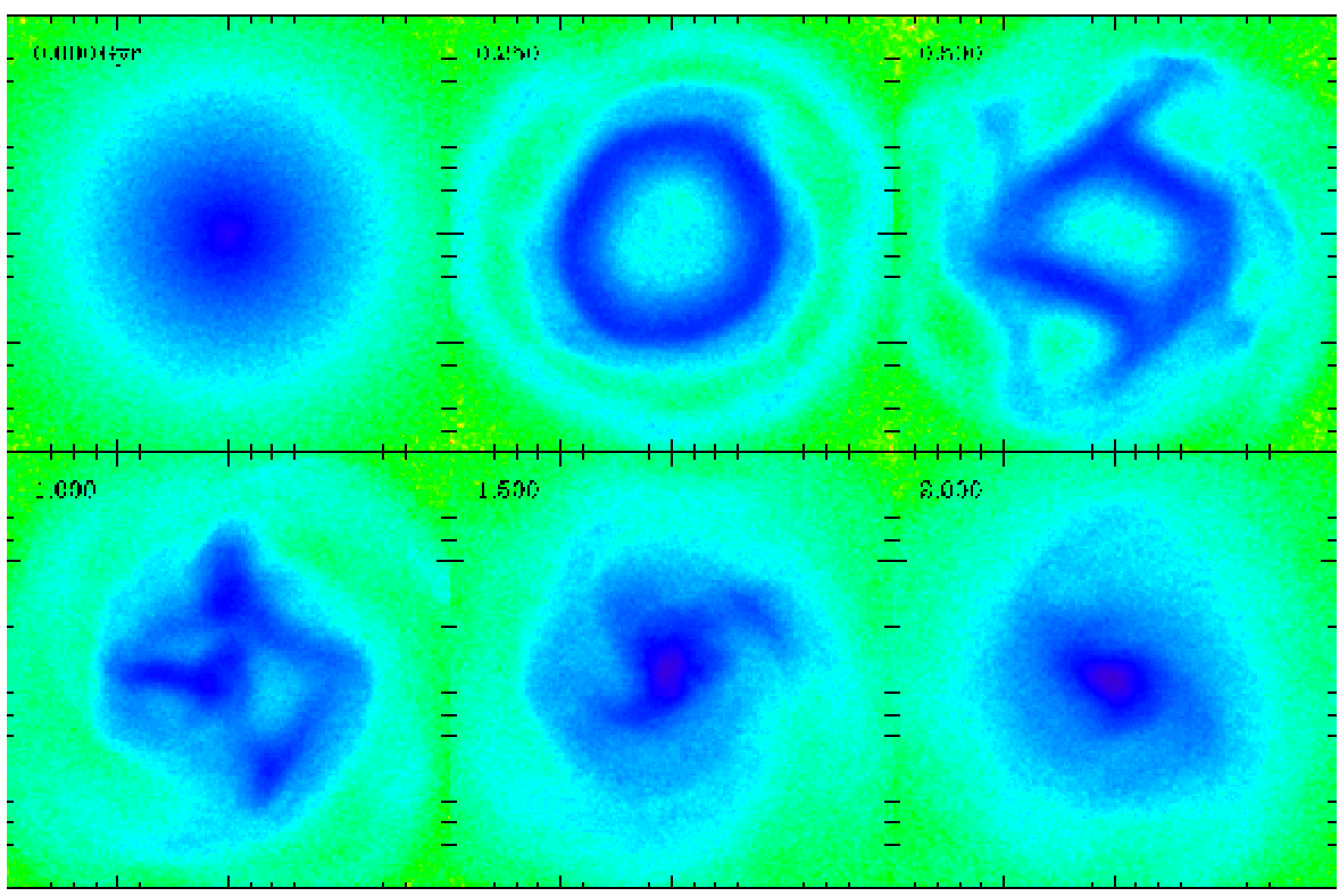}
\caption{Projected Stellar Density images for the SpNGNB simulation out to 2 Gyr.  The length of the box is 10 kpc.}
\end{center}
\end{figure}

\begin{figure}
\begin{center}
\includegraphics[scale=0.4]{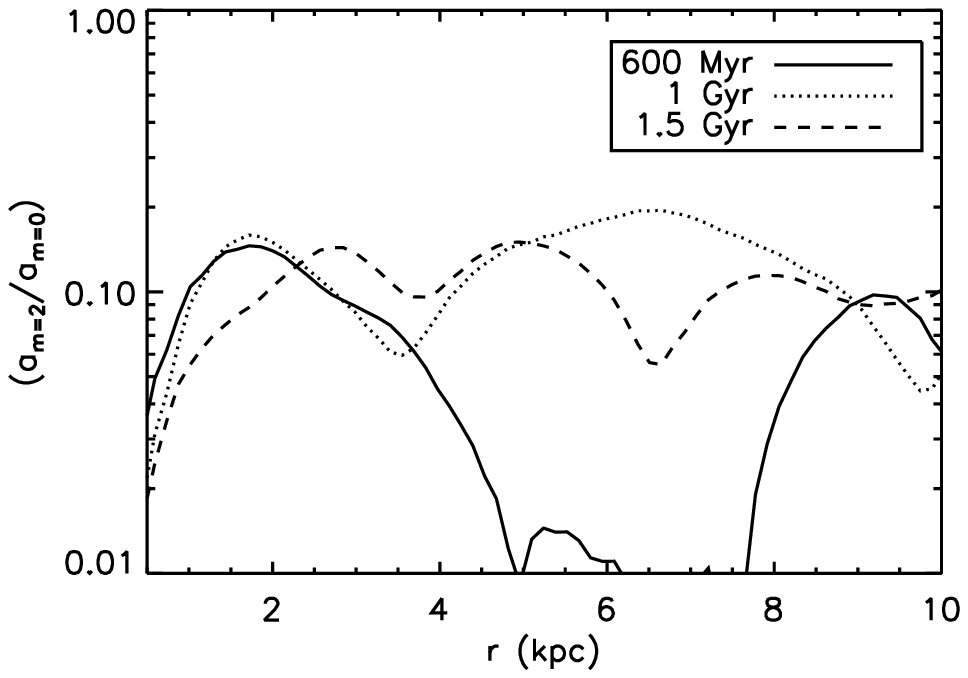}
\caption{The $m=2$ Fourier component of the projected stellar density at three time snaps for the SpNGNB simulation.}
\end{center}
\end{figure}

\begin{figure}
\begin{center}
   \includegraphics[scale=0.4]{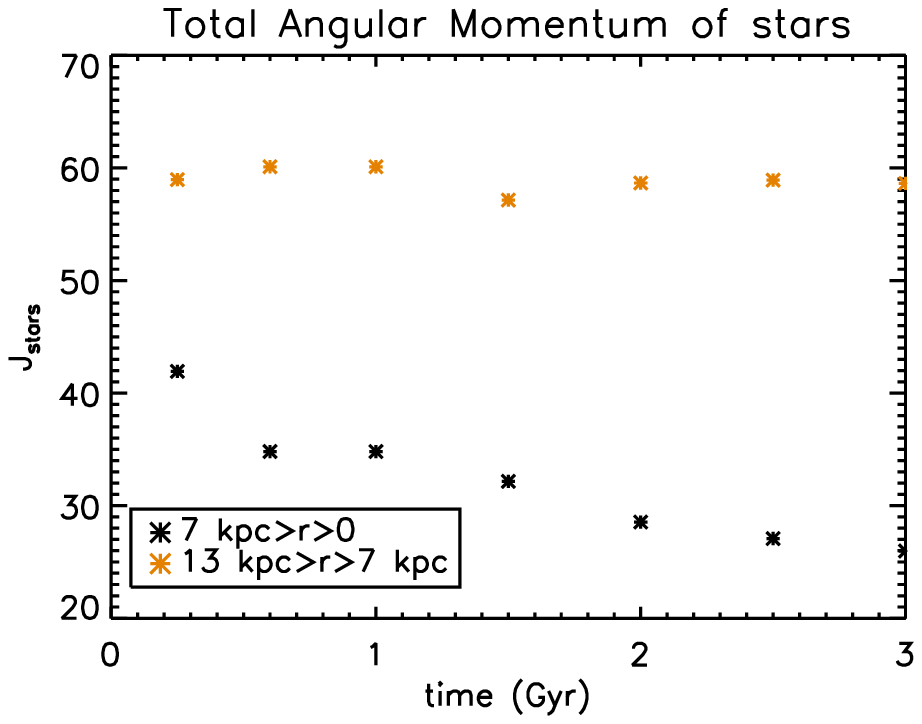}
   \includegraphics[scale=0.4]{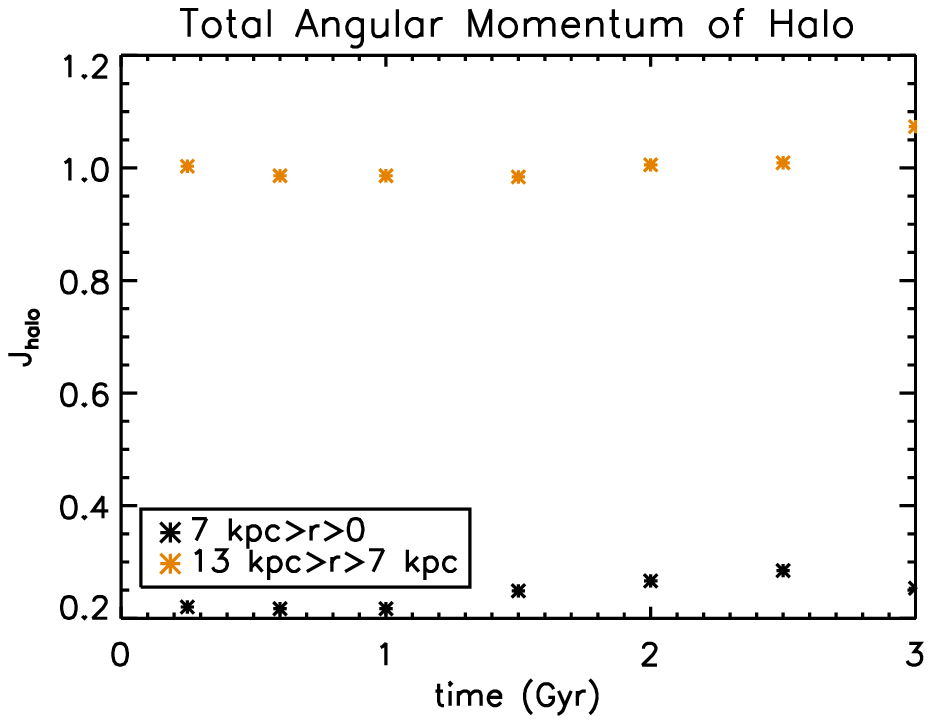}
   \caption{The time evolution of $J_{z}$ for the stars (a) and the halo (b), for the SpNGNB simulation for $0<r<7~\rm kpc$ and $7~\rm kpc<r<13~\rm kpc$}
\end{center}
\end{figure}

\begin{figure}
\begin{center}
\includegraphics[scale=0.4]{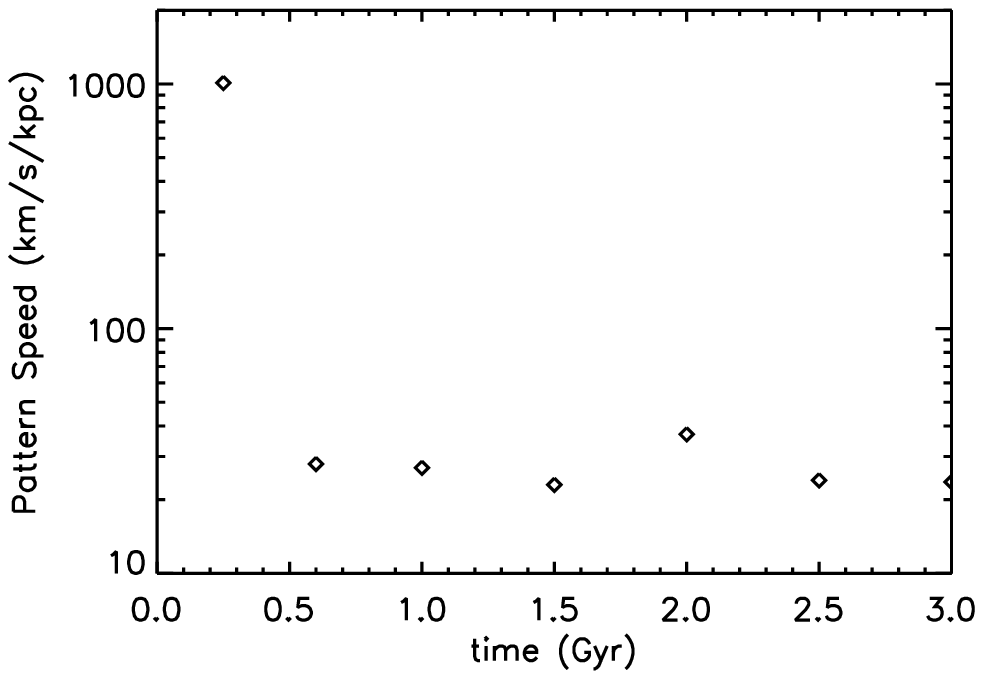}
\caption{The time variation of the pattern speed, out to 3 Gyr (identified using Tremaine \& Weinberg 1984) for the SpNGNB simulation (no gas, no bulge, stars and live DM halo only)}
\end{center}
\end{figure}

We depict in Figure 1 the projected (i.e., we take our stellar density in x,y,z and project it on the z-axis) stellar density ($\Sigma$) out to 2 Gyr for the SpNGNB simulation.  This stellar disk is only somewhat susceptible to spiral structure as it is initialized with $Q \ga 1$ in the inner parts of the disk, with $Q \sim 1$ in the outer regions.  A more quantitative representation of the spiral structure is given in Figure 2, where we have plotted the $m=2$ Fourier component of the stellar surface density.  We obtain this by taking the amplitude of the real and imaginary components of $a_{m=2}(r)$, where $a_{m}(r,t)$ is given by:
\begin{equation}
a_{m}(r,t)=\frac{1}{2\pi}\int_{0}^{2\pi} \Sigma(r,\phi,t)e^{-im\phi}d\phi  \;.
\end{equation}
Figure 2 shows that that the power in the $m=2$ disturbance (relative to the m=0 axisymmetric mode) varies by a factor of $\sim 10$ in magnitude peak-to-peak in the inner regions of the disk ($r \la 8~\rm kpc$) over the timescale of a Gyr.  

The spiral patterns disappear within an orbital time due to disk heating, an effect also seen by SC84.  Rybicki (1971) estimated that the relaxation timescale is comparable to the orbital time for purely stellar razor-thin disks.  The swing amplification mechanism (Toomre 1981) is effective for low values of $Q$ - as the spiral patterns swing from leading to trailing in a shearing disk, they are amplified across co-rotation.  In the absence of dissipation, this increases the velocity dispersion of the stars secularly until the disk is no longer susceptible to this amplification mechanism.

As noted earlier, the total angular momentum of the simulations is conserved.  However, different galaxy components are free to exchange angular momenta.  We show in Figure 3a the angular momentum (in arbitrary units) of the stars and dark matter halo particles within $0> r > 7~\rm kpc$ and $7~\rm kpc > r > 13~\rm kpc$ (Figure 3b).  The stars inside co-rotation (which is at $\sim~7~\rm~kpc$) lose angular momentum to the halo; of particular note is the steady decline of angular momentum in the stellar disk.  The massive slowly rotating halo is then a net sink of angular momentum, as has been found in previous simulations of bar instabilities with live halos (Athanassoula 2003; Martinez-Valpuesta et al. 2006; Berentzen et al. 2007).  The basic idea that energy and angular momentum can be absorbed by the outer part of a galaxy if it is rotating slower than a given mode was discussed in the early studies of angular momentum transport in galaxies (Lynden-Bell \& Kalnajs 1972).

We show in Figure 4 the time variation of the pattern speed, computed following Tremaine \& Weinberg (1984).   This method derives the pattern speed on the assumption of a well-defined pattern speed and the tracer ($\Sigma$) obeying the continuity equation.  This gives a pattern speed in terms of the projected surface densities ($\Sigma(x,z)$), which we have computed in the center of mass frame, and velocities ($v(x,z)$) of the stars:
\begin{equation}
\Omega_{p}=\frac{\int_{-\infty}^{\infty} h(z)dz \int_{-\infty}^{\infty} \Sigma(x,z) v(x,z) dx}{\int_{-\infty}^{\infty} h(z)dz \int_{-\infty}^{\infty} \Sigma(x,z) x dx} \;,
\end{equation}
which we have expressed the pattern speed here for an edge-on disk; the weight function $h(z)dz$ can be chosen arbitrarily.  For this simulation, we find that the pattern speed is on average $\sim 27~\rm km/s/kpc$ (for $t \ga 500~\rm Myr$), which places co-rotation at $\sim 7~\rm kpc$, and is in agreement with expectations from modal theory as we discuss later.  The early-time ($t=250~\rm Myr$) pattern speed is high (which would imply a co-rotation radius close to the very inner regions of the disk), as the $m=2$ spiral mode is not yet the dominant mode in most of the disk.  For this collisionless simulation, there is a 97 \% change in the pattern speed out to 3 Gyr, and a 38 \% change for $t \geq 500~\rm Myr$.  As we discuss later, simulations that include a gaseous component show less variance in their pattern speeds as a function of time.

\section{Self-Excited Quasi-Steady Spiral Structure - Simulations with Dark Matter, Stars, and Gas}

We now proceed to describing our results from simulations that include a gaseous component.  We focus our analysis on angular momentum transport from the gas to the stars.  The torque exerted by the self-gravitating gas disk on the stellar disk is given by - 
\begin{equation}
\tau_{\star}=\int_{0}^{2\pi}\int_{0}^{\infty}\Sigma_{\star}(r,\phi)\frac{\partial\Phi_{\rm gas}(r,\phi)}{\partial\phi}rdrd\phi \;,
\end{equation}
where it is specifically the asymmetric components that contribute to this integral.  

\begin{figure}
\begin{center}
\includegraphics[scale=0.25]{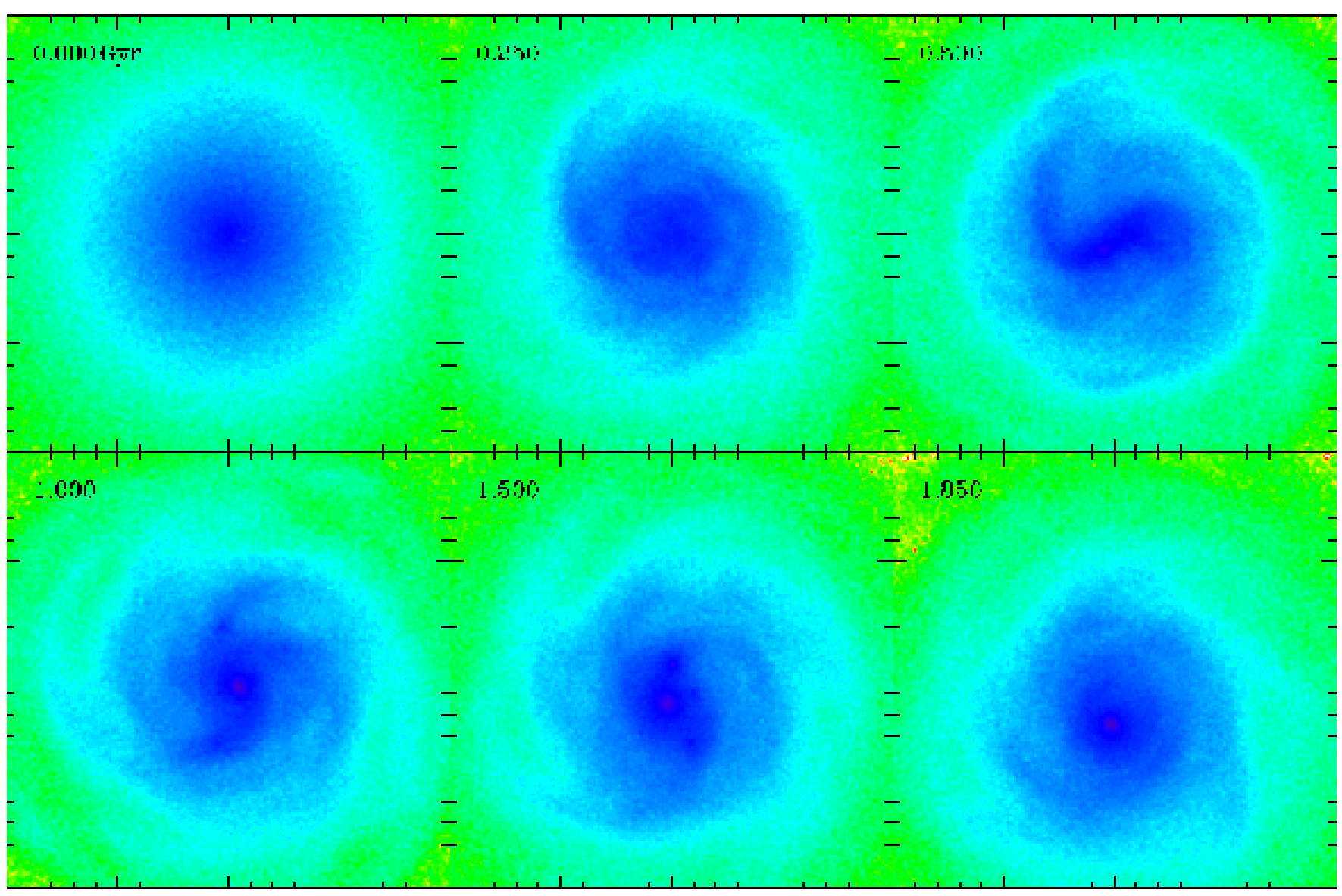}
\caption{Projected Stellar Density images for the SpGNSFR simulation out to 2 Gyr.  The length of the box is 10 kpc.}
\end{center}
\end{figure}

\begin{figure}
\begin{center}
\includegraphics[scale=0.25]{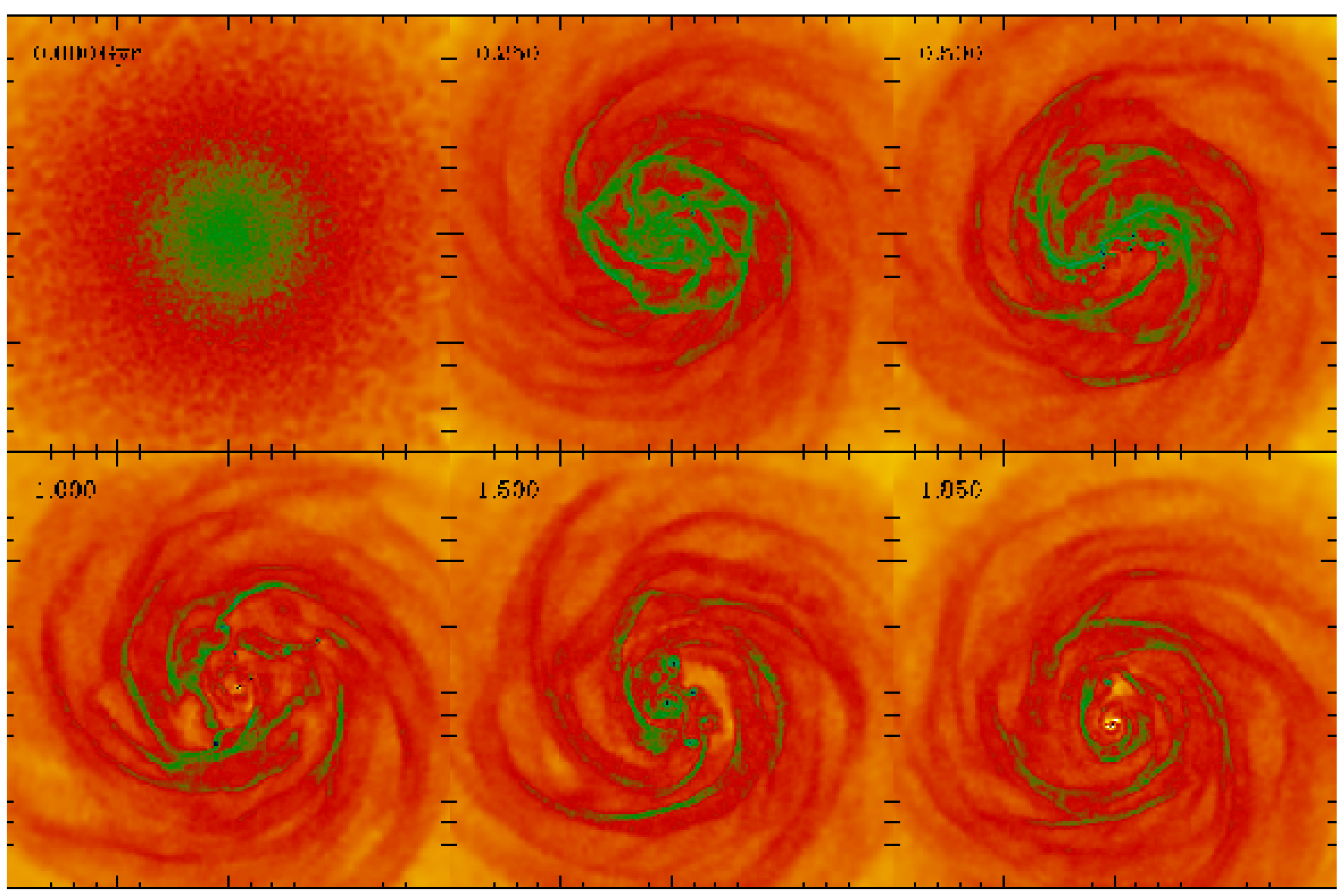}
\caption{Projected Gas Density images for the SpGNSFR simulation out to 2 Gyr.  The length of the box is 10 kpc.}
\end{center}
\end{figure}

We depict in Figure 5 the projected stellar density from the SpGNSFR simulation, which contains 10\% of the disk mass in the gaseous component.  Star formation is turned off in this simulation.  As is clear, the stellar disk is unstable to spiral structure for a longer time when there is gas, and $\it{thin~spidery~asymmetric}$ spiral arms are seen in the gaseous component (Figure 6).  Figure 7 shows the time variation of the power in the $m=2$ Fourier mode, and can be contrasted with Figure 2. In this case, when gas is included in the simulation, the $m=2$ disturbance is roughly time-steady - there is less than a factor of three varation in the power peak-to-peak.

\begin{figure}
\begin{center}
\includegraphics[scale=0.4]{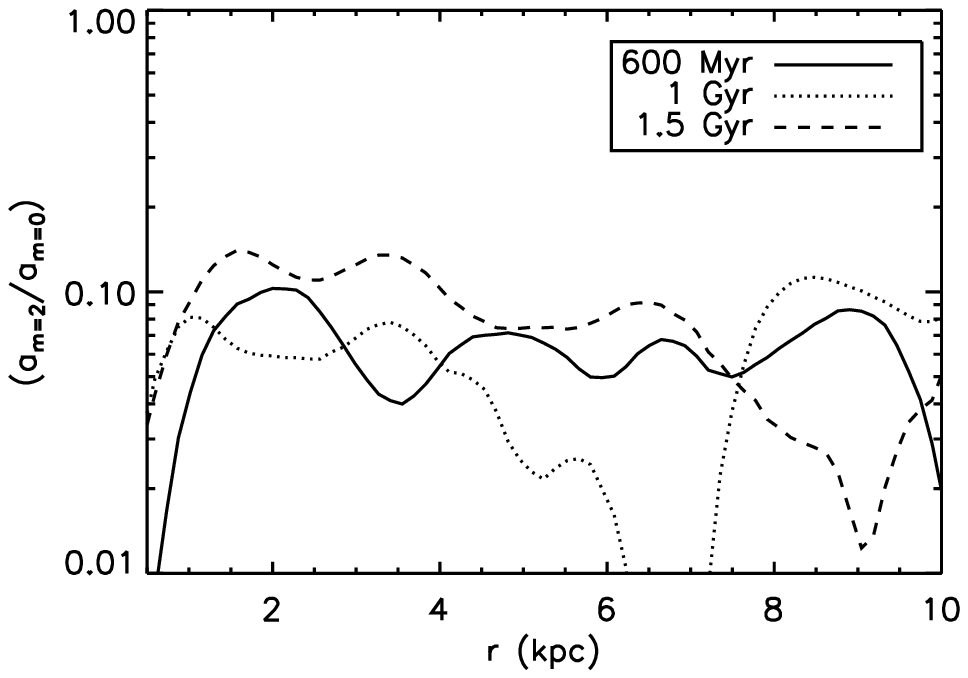}
\caption{The $m=2$ Fourier component of the projected stellar density at three time snaps for the SpGNSFR simulation.}
\end{center}
\end{figure}

\begin{figure}
\begin{center}
   \includegraphics[scale=0.4]{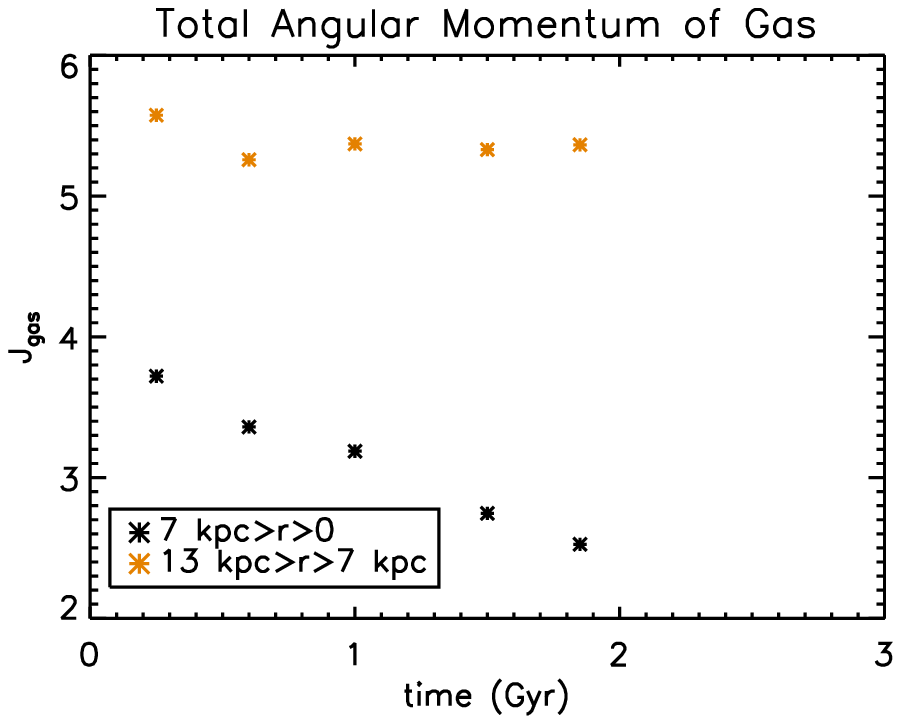}
   \includegraphics[scale=0.4]{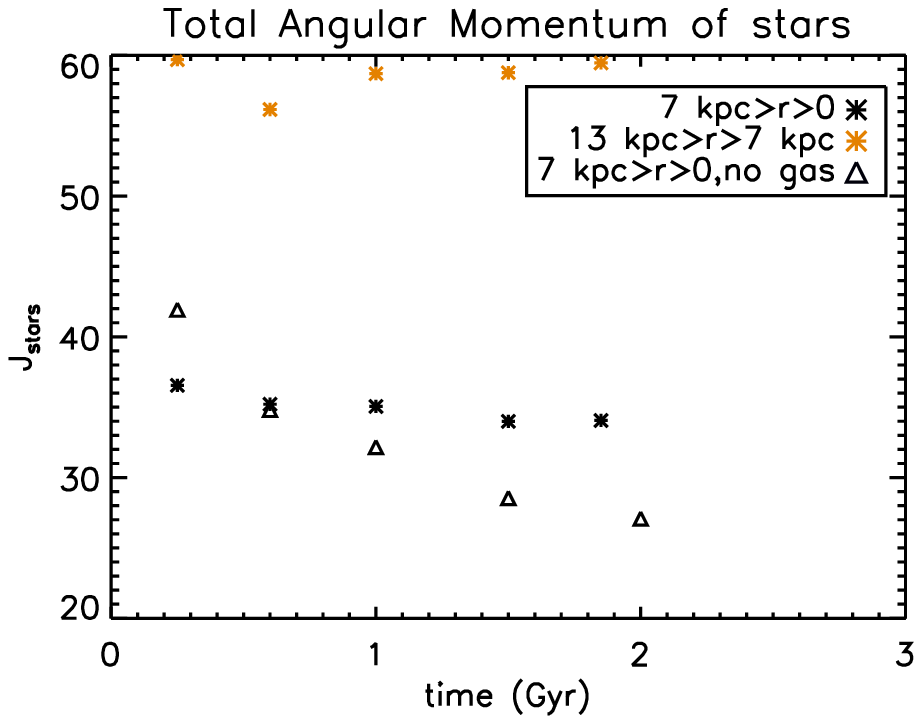}
   \includegraphics[scale=0.4]{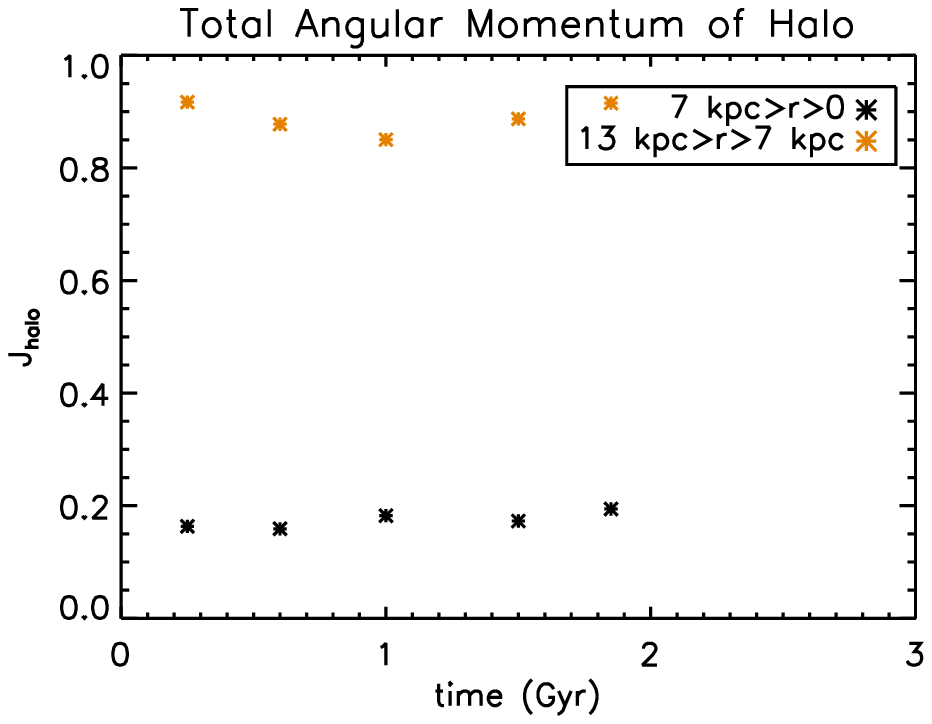} 
   \caption{The time evolution of $J_{z}$ for (a) the gas, (b) the stars; overplotted is $J_{z}$ for the stars for the SpNGNB simulation as a reference point for a case without gas (in this case the stars lose 7 \% of their angular momentum inside of $7~\rm kpc$ out to 2 Gyr in contrast to the 35 \% loss in the purely collisionless case), and (c) the halo for the SpGNSFR simulation within $0<r<7~\rm kpc$ and $7~\rm kpc<r<13~\rm kpc$}
\end{center}
\end{figure}

Figure 8 depicts the evolution of angular momentum for the gas, stars, and halo interior and exterior to co-rotation.  Angular momentum transport from the gas to the stars interior to co-rotation is able to torque the stellar disk and serves to self-regulate the disk structure.  Figure 8b shows the angular momentum of the stars is considerably more time-steady than its purely collisionless analogue (Figure 3a), as the stars are gaining the angular momentum that is lost by the gas (Figure 8a).  There continues to be a slight net transfer to the live halo.  

The sharp azimuthal response of the gas and the smooth response of the stars is shown in Figure 9 for the 250 Myr snapshot - this also shows that the gas leads the stars, with a phase shift of $\sim 0.5~\rm radian$ close to co-rotation.  Given that the potential traces the density, Figure 9 also shows why the gas is able to exert torques on the stars -- $\frac{d}{d\phi}\Phi_{\rm gas} \gg \frac{d}{d\phi}\Phi_{\rm stars}$, as the gas is compressed and shocks, it forms the sharp azimuthal features shown in Figure 9.  The response of the gas to a time-steady forcing potential in the nonlinear regime was studied by Shu, Milione \& Roberts (1973) (henceforth SMR).  SMR found that the response of the gas can be quite nonlinear even when the amplitude of the stellar spiral forcing is a small fraction of the axisymmetric field.  The response of the gas is much sharper than that of the stars as the response of some galactic component to a forcing potential is proportional to the inverse square of its dispersive speed.  SMR also discussed in detail how the breadth of the zone of high gas compression depends on whether the Doppler shifted phase velocity of the stellar density wave (or the component of the circular motion that is perpendicular to the wave front) is greater or less than the sound speed of the gaseous component, which results in narrow and broad zones of gas compression respectively.  In this simulation, where star formation is turned off, there is no energy injection from supernovae.  This allows the sound speed of the gas to be low enough to satisfy the criterion of SMR, and can be contrasted with the cases discussed later in this paper where star formation is turned on along with an injection of energy prescribed by the model of SH03 which results in broad gaseous arms, particularly in the outer parts of the galaxy.  Time-dependent numerical calculations by CLS of the response of the gaseous disk when forced by a spiral stellar potential displayed similar sharp azimuthal features. 

\begin{figure}
\begin{center}
\includegraphics[scale=0.4]{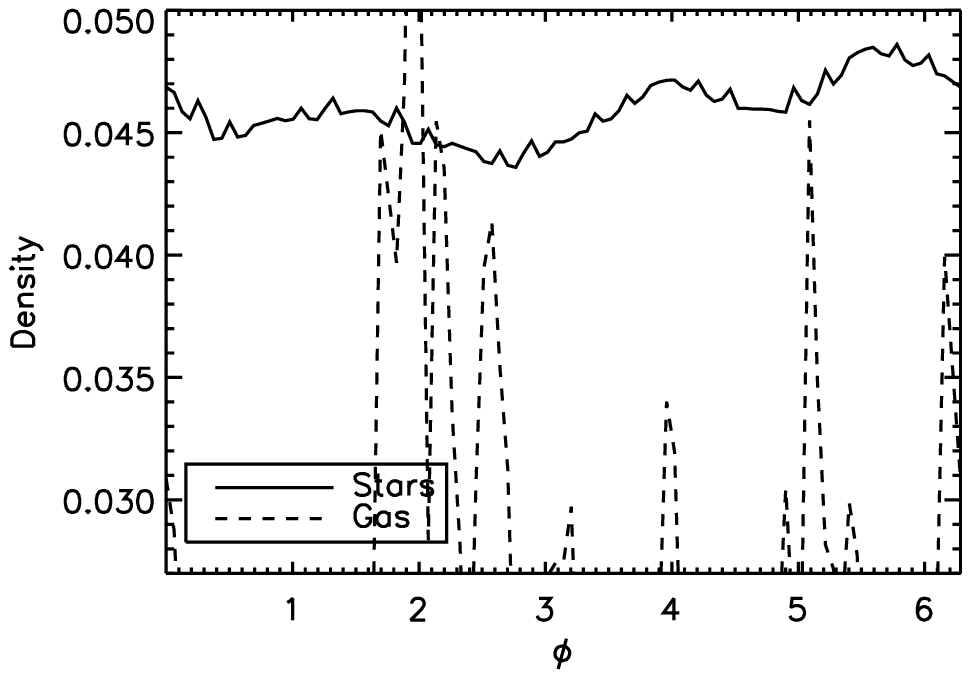}
\caption{Azimuthal phase shift in density (arbitrarily scaled) between the gas and the stars in the SpGNSFR simulation at 250 Myr at co-rotation ($r=7~\rm kpc$).}
\end{center}
\end{figure}

\begin{figure}
\begin{center}
\includegraphics[scale=0.4]{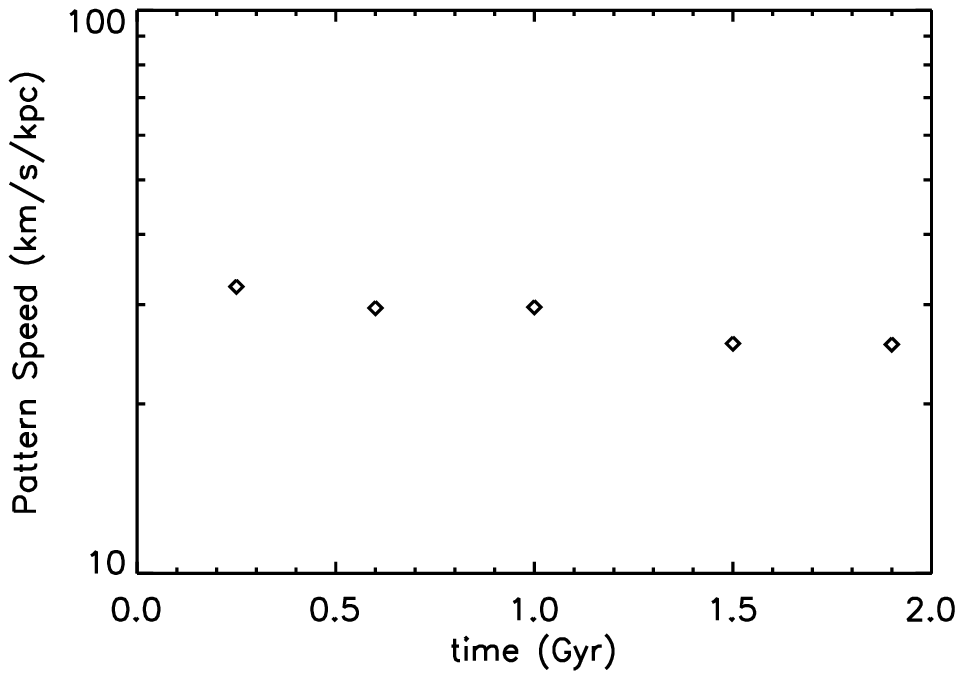}
\caption{The time variation of the pattern speed, out to 2 Gyr (identified using Tremaine \& Weinberg 1984) for the SpGNSFR simulation, with star formation turned off.}
\end{center}
\end{figure}

Figure 10 displays the time variation of the pattern speed for the SpGNSFR simulation.  The average pattern speed of this simulation is $\sim 28~\rm km/s/kpc$, which places co-rotation at $r\sim 7~\rm kpc$, and is in agreement with analytic expectations of the pattern speed from modal theory, i.e., a few scale lengths (Bertin et al. 1998) and from observational comparisons (Lin et al. 1969; Shu et al. 1970).  There is no inner Lindblad resonance, and the outer Lindblad resonance is at $\sim 15~\rm kpc$.  We note that in contrast to the SpNGNB collisionless simulation, this simulation exhibits less variance in its pattern speed, with a maximum difference of 21 \% out to $\sim 2~\rm Gyr$.

It is interesting to note that the average radial velocities of the gas (at times when the
spiral structure is well established) are of order $\sim 30$\% of the sound speed, which would cause gas to flow radially inwards $\sim 2~\rm kpc$ on the time scale of a Gyr.  This
effect cannot be neglected on the scale of the spiral arms, and the radial velocities in these simulations are in rough agreement with observational estimates (Wong et al. 2004).  Radial inflow may also be produced in
the outskirts of isolated galaxies due to the torque on the gas from new stars that exist along spiral
arms far outside the optical radius as seen in recent Galex images (Thilker et al. 2007).  
We investigate the large scale transport of gas from extended HI disks as a possible mechanism for 
fueling star formation over many Gyr in a forthcoming paper. 

\subsection{Star Formation Cases}

\begin{figure}
\begin{center}
\includegraphics[scale=0.25]{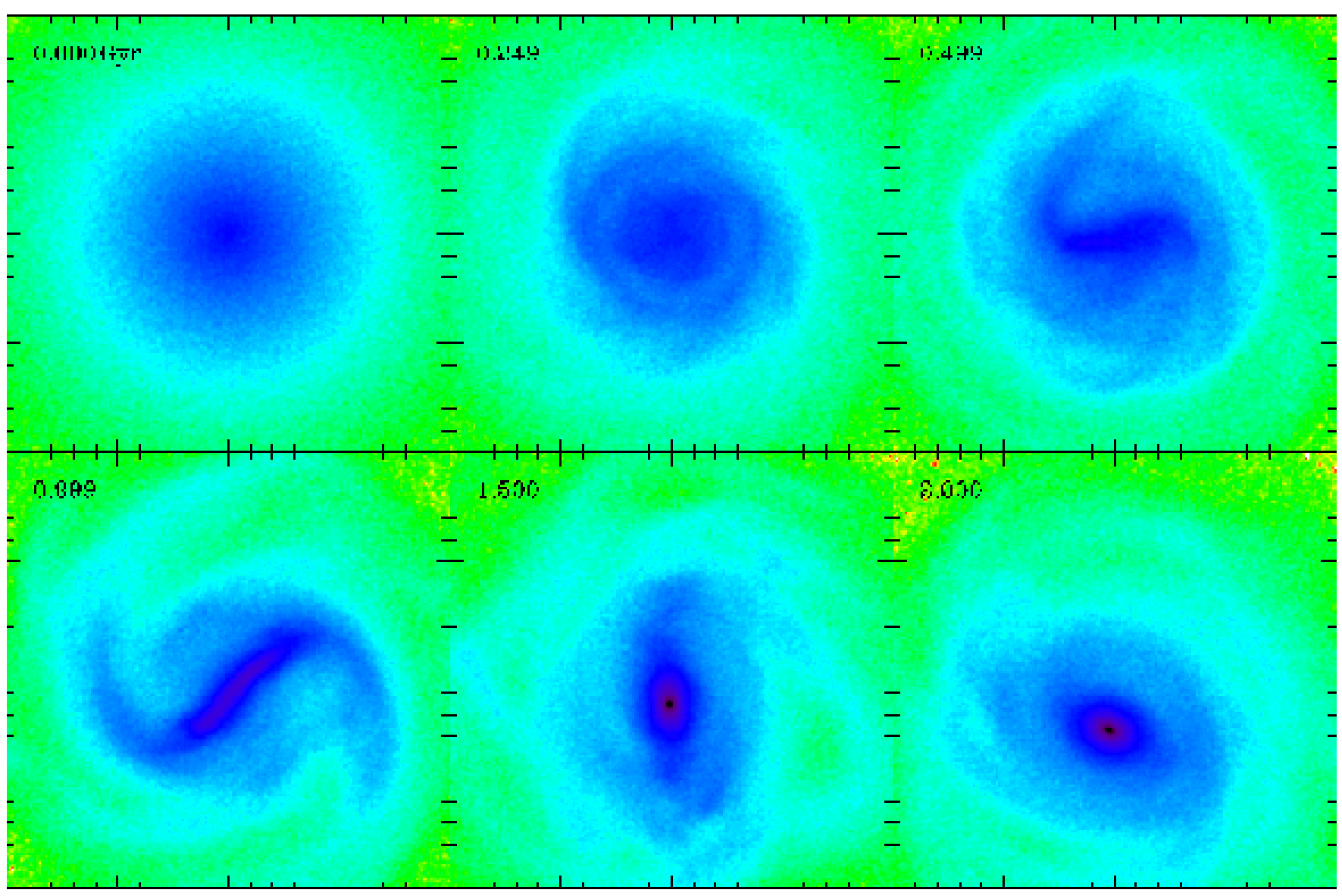}
\caption{Projected Stellar Density images for the SpGKS simulation out to 2 Gyr.  The length of the box is 10 kpc.}
\end{center}
\end{figure}

\begin{figure}
\begin{center}
\includegraphics[scale=0.25]{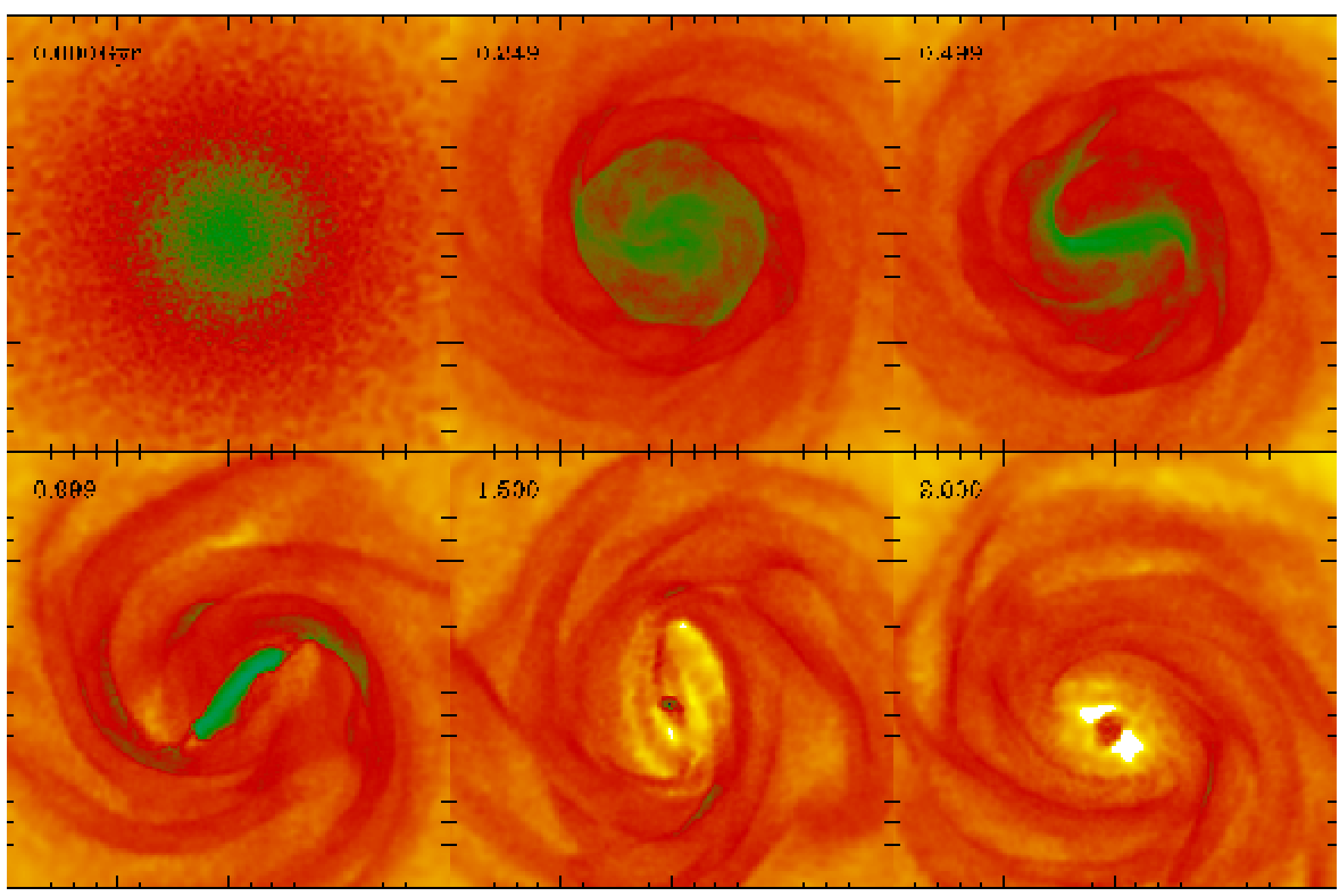}
\caption{Projected Gas Density images for the SpGKS simulation out to 2 Gyr.  The length of the box is 10 kpc.}
\end{center}
\end{figure}

\begin{figure}
\begin{center}
\includegraphics[scale=0.25]{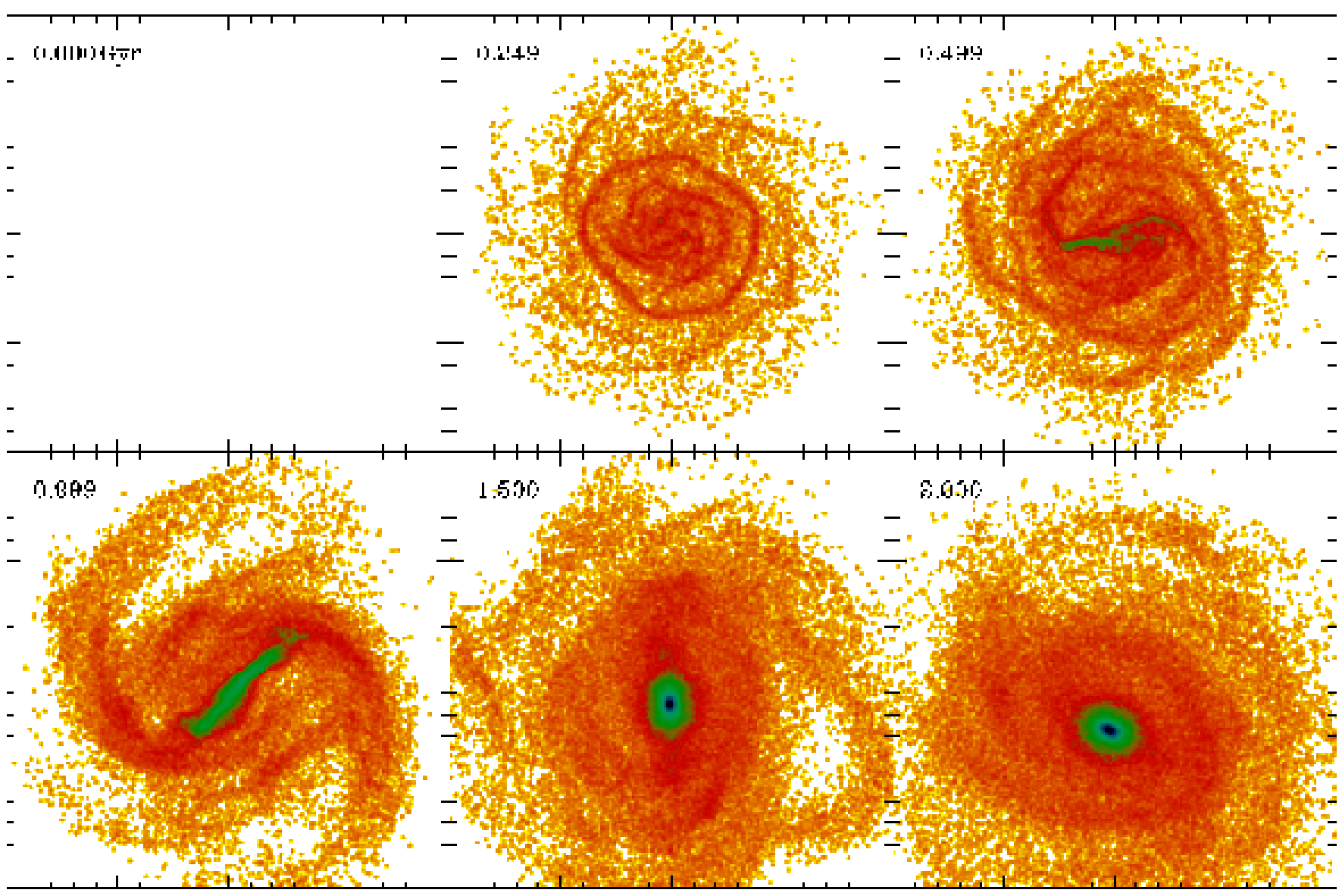}
\caption{Projected density for the new stars formed in the SpGKS simulation out to 2 Gyr.  The length of the box is 10 kpc.  The first snapshot is blank as there is no star formation at $t=0$.}
\end{center}
\end{figure}

Figures 11 and 12 show the projected stellar and gas surface densities for the SpGKS simulation, where star formation is prescribed via the Kennicutt-Schmidt relation with a gas consumption time scale of 4.5 Gyr.  Although the results are qualitatively similar to the SpGNSFR case, the spiral structure in the gas is not as sharp due to energy injection from star formation as prescribed in the multiphase model of SH03 (which may be expected from the work of SMR as we have discussed in \S 4), and the stellar disk is also not as unstable to spiral structure over 2 Gyr.  Figure 14 shows the time evolution of the $m=2$ Fourier component of the stellar surface density, which is also in qualitative agreement with the SpGNSFR simulation, although the variation in power over a Gyr is somewhat larger.  The time evolution of the angular momentum (Figures 15) of the stars is not as time steady as in the SpGNSFR simulation - over the period of 2 Gyr, the angular momentum of the stars declines by about 20 \%, in contrast to the 7 \% decrease for the SpGNSFR simulation, but both of these are below the 30\% stellar angular momentum loss seen in the SpNGNB simulation that did not include gas.  Of note is the projected density of the new stars as shown in Figure 13, which is more akin to the sharper response of the gas shown in Figure 12.  Figure 16 depicts the density vs azimuth at 7 kpc for the SpGKS simulation and is similar to the SpGNSFR case, though with a slightly broader azimuthal response in the gaseous component.  The time variation of the pattern speed is shown in Figure 17, with an average pattern speed of $\sim 28~\rm~km/s/kpc$, which places co-rotation at $\sim~7~\rm kpc$.  There is a maximum variance of 27 \% in the pattern speed for the SpGKS simulation out to 3 Gyr.

\begin{figure}
\begin{center}
\includegraphics[scale=0.4]{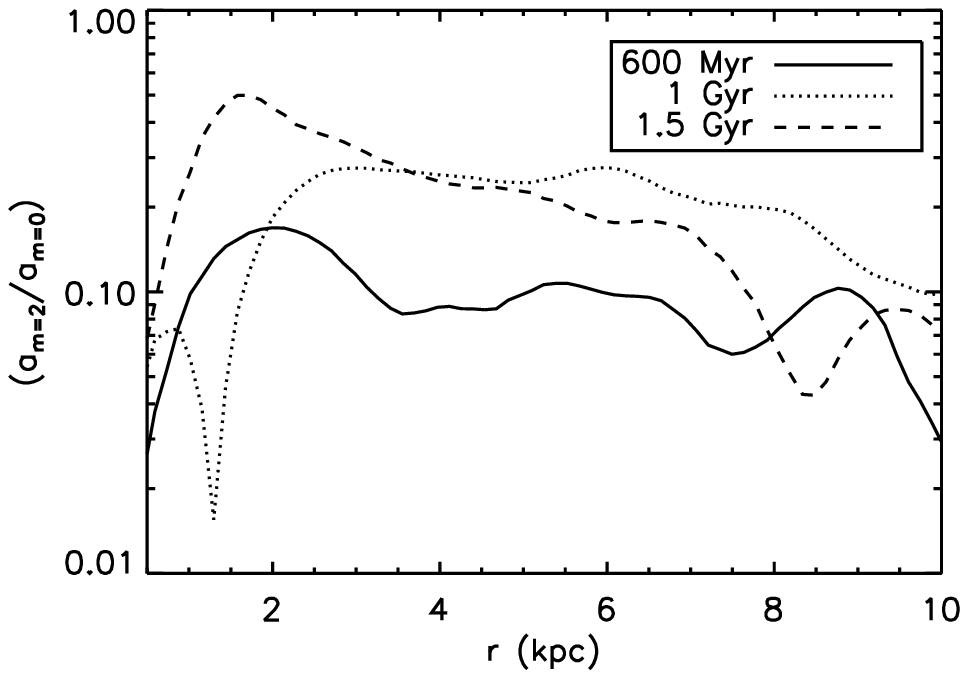}
\caption{The $m=2$ Fourier component of the projected stellar density at three time snaps for the SpGKS simulation.}
\end{center}
\end{figure}

\begin{figure}
\begin{center}
   \includegraphics[scale=0.4]{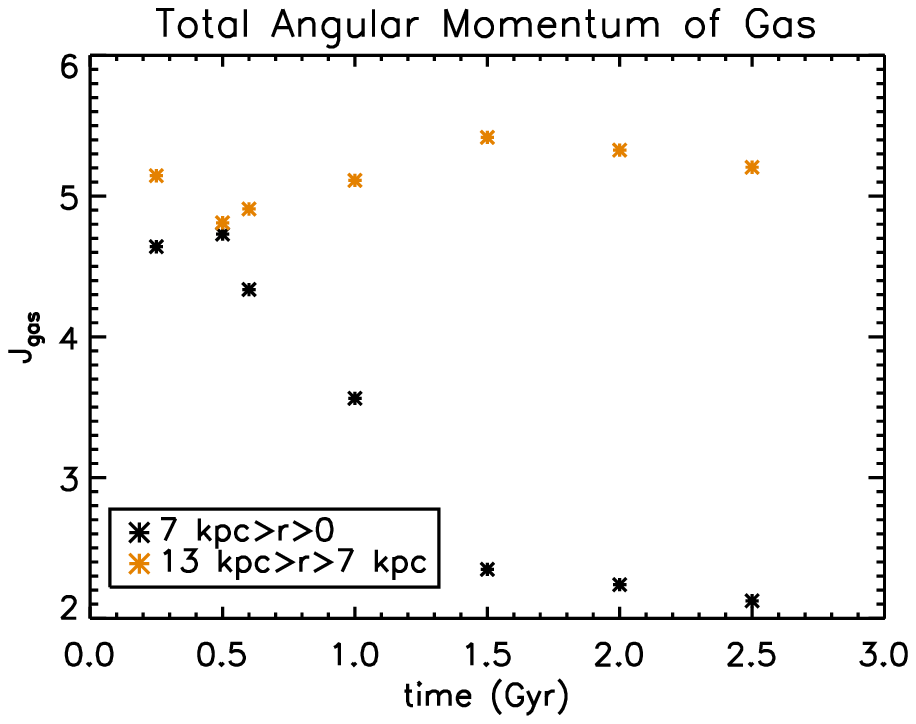}
   \includegraphics[scale=0.4]{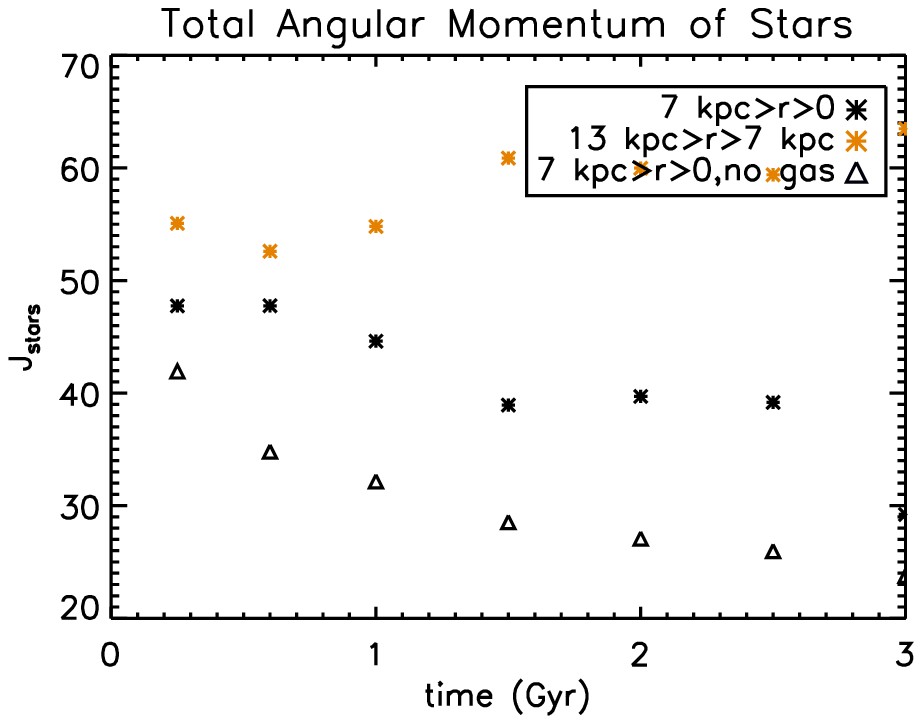}
   \includegraphics[scale=0.4]{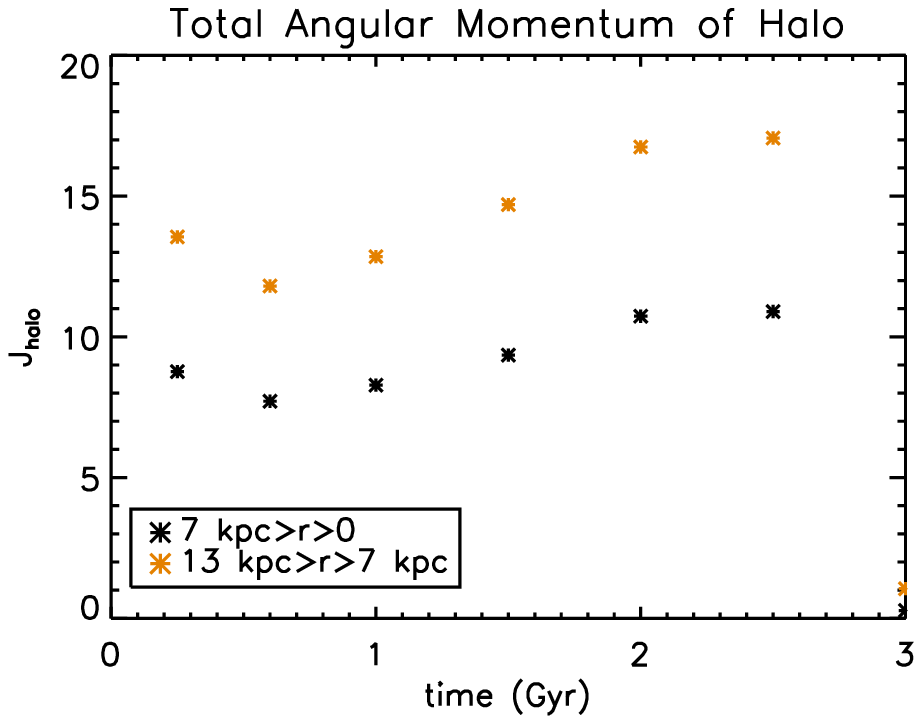} 
   \caption{The time evolution of $J_{z}$ for (a) the gas, (b) the stars shown along with the comparison to the purely collisionless SpNGNB case (in this case the stars lose 38 \% of their angular momentum inside of $7~\rm kpc$ out to 3 Gyr in contrast to the 43 \% loss seen in the purely collisionless case), and (c) the halo for the SpGKS (in my notation SpGKS) simulation within $0<r<7~\rm kpc$ and $7~\rm kpc<r<13~\rm kpc$}
\end{center}
\end{figure}

\begin{figure}
\begin{center}
\includegraphics[scale=0.4]{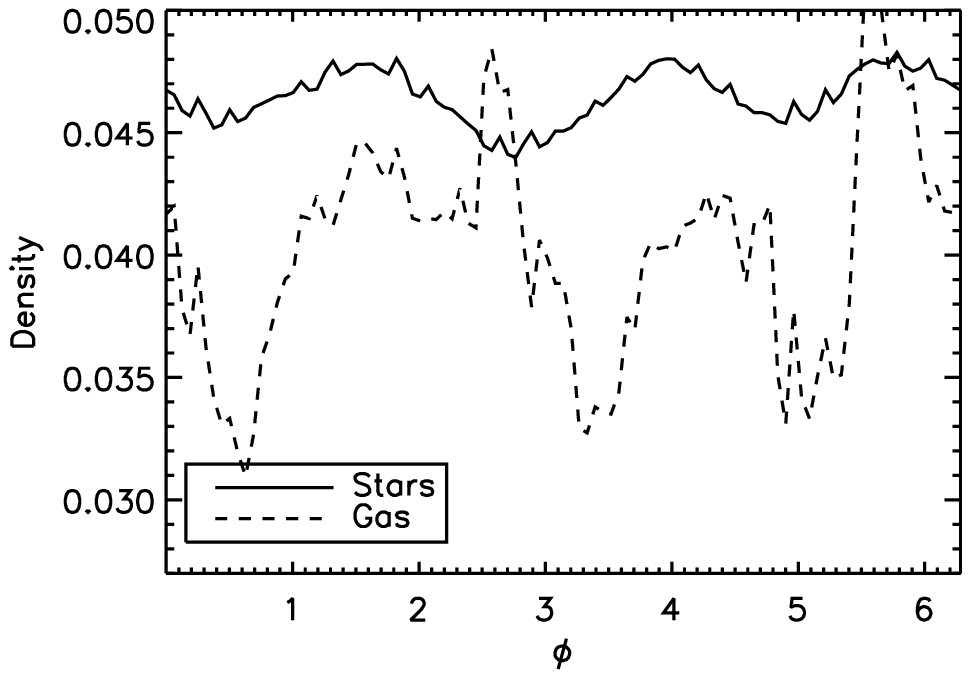}
\caption{Azimuthal phase shift in density (arbitrarily scaled) between the gas and the stars in the SpGKS simulation at co-rotation at 250 Myr.}
\end{center}
\end{figure}

\begin{figure}
\begin{center}
\includegraphics[scale=0.4]{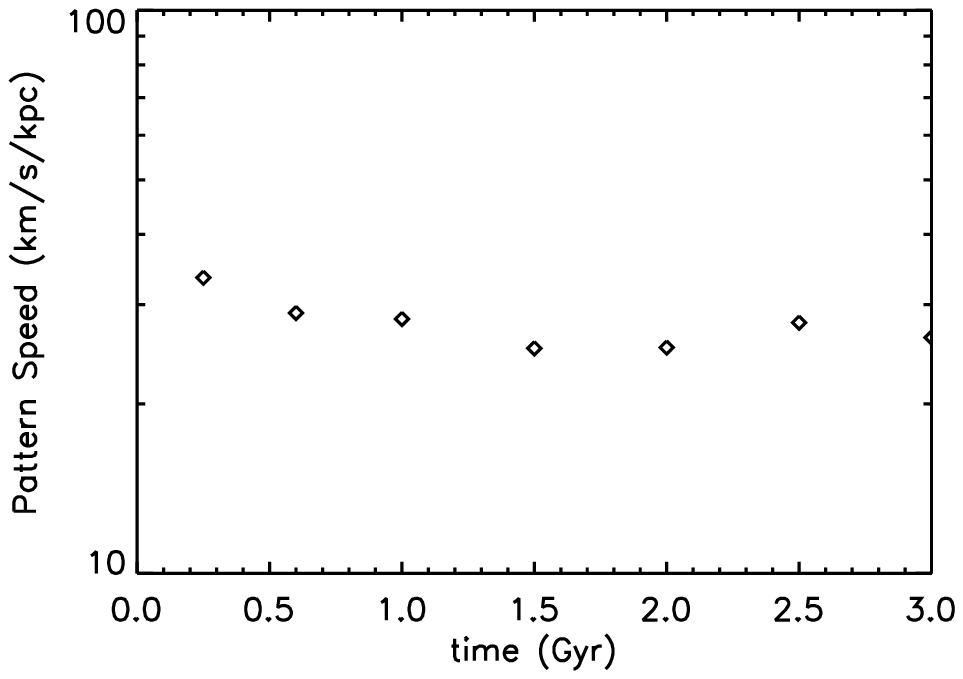}
\caption{The time variation of the pattern speed, out to 3 Gyr (identified using Tremaine \& Weinberg 1984) for the SpGKS simulation.}
\end{center}
\end{figure}

Gas consumption time scales in nearby spiral galaxies are inferred to be on the order of a few Gyr (Kennicutt 1983; Wong \& Blitz 2002; Crosthwaite \& Turner 2007), with a range between 1-10 Gyr (Wong \& Blitz 2002).  The values are uncertain both due to use of various tracers to infer the rate of star formation and hence the gas depletion time scales, as well as due to the necessary use of globally averaged measures which recent works have been able to address observationally (see e.g., Kennicutt et al. 2007).  Our use of a density dependent star formation prescription is simplistic; star formation in real galaxies may be mediated magnetically (Shu et al. 2007) and or through turbulence (Krumholz \& McKee 2005), and may be understood in a simple way in terms of the pressure (Blitz \& Rosolowsky 2006).  Nonetheless, use of this standard star formation prescription is fairly pervasive (SH03; Springel et al. 2005; Cox et al. 2006) and does agree with globally averaged measures of the star formation rate in a diverse range of galaxy types.   

\begin{figure}
\begin{center}
\includegraphics[scale=0.25]{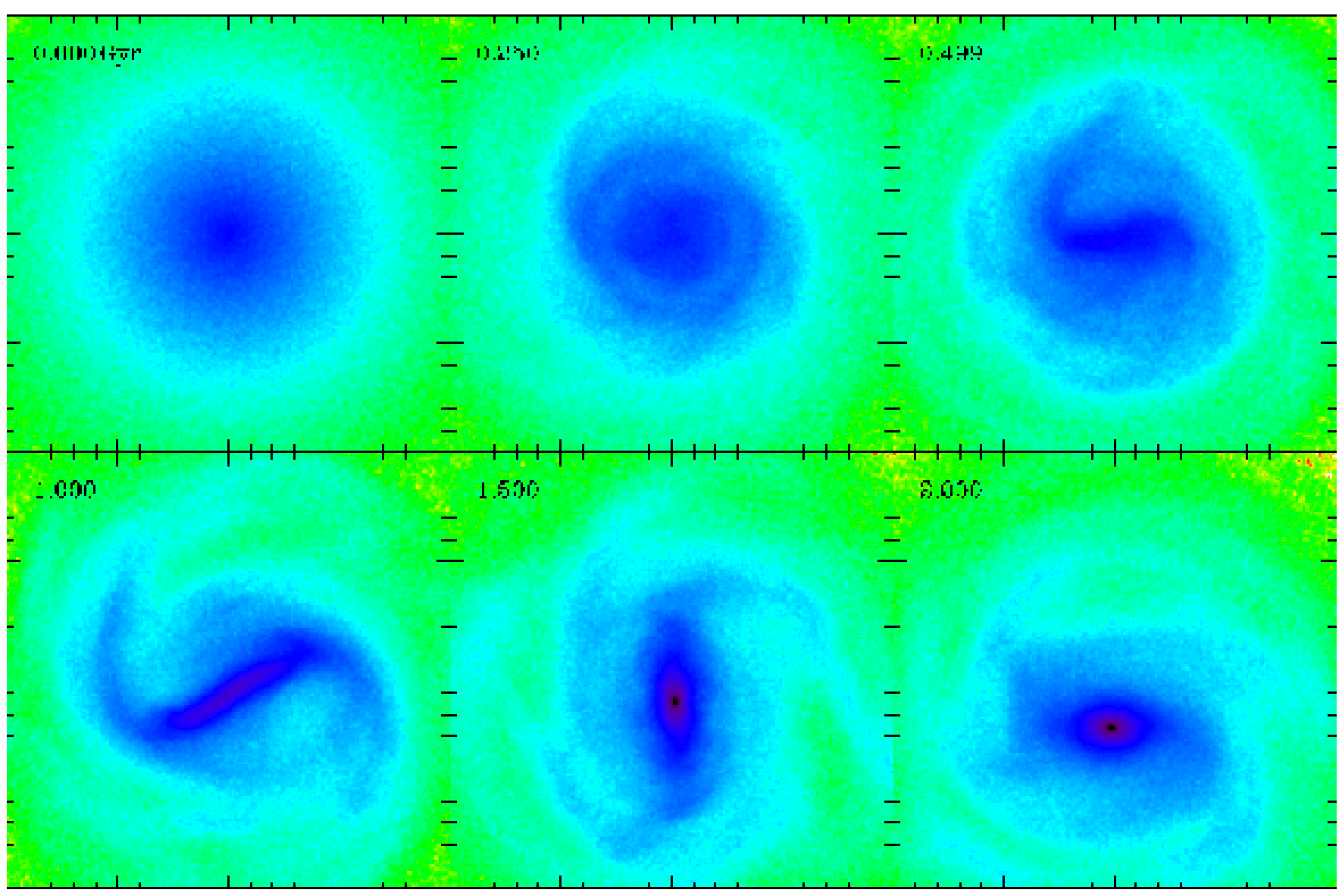}
\caption{Projected Stellar Density images for the SpGKS8 simulation out to 2 Gyr.  The length of the box is 10 kpc.}
\end{center}
\end{figure}

\begin{figure}
\begin{center}
\includegraphics[scale=0.25]{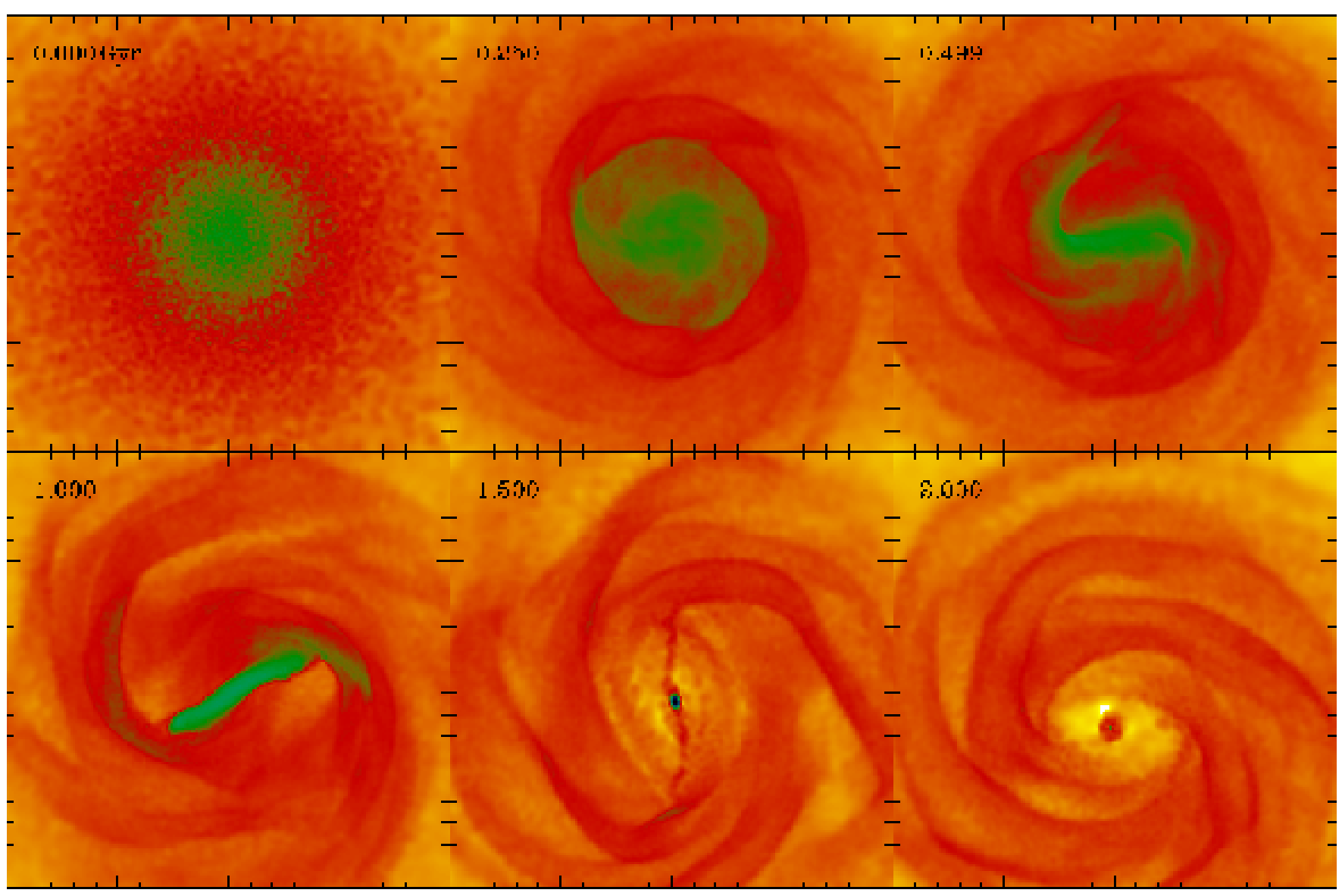}
\caption{Projected Gas Density images for the SpGKS8 simulation out to 2 Gyr.  The length of the box is 10 kpc.}
\end{center}
\end{figure}

\begin{figure}
\begin{center}
\includegraphics[scale=0.25]{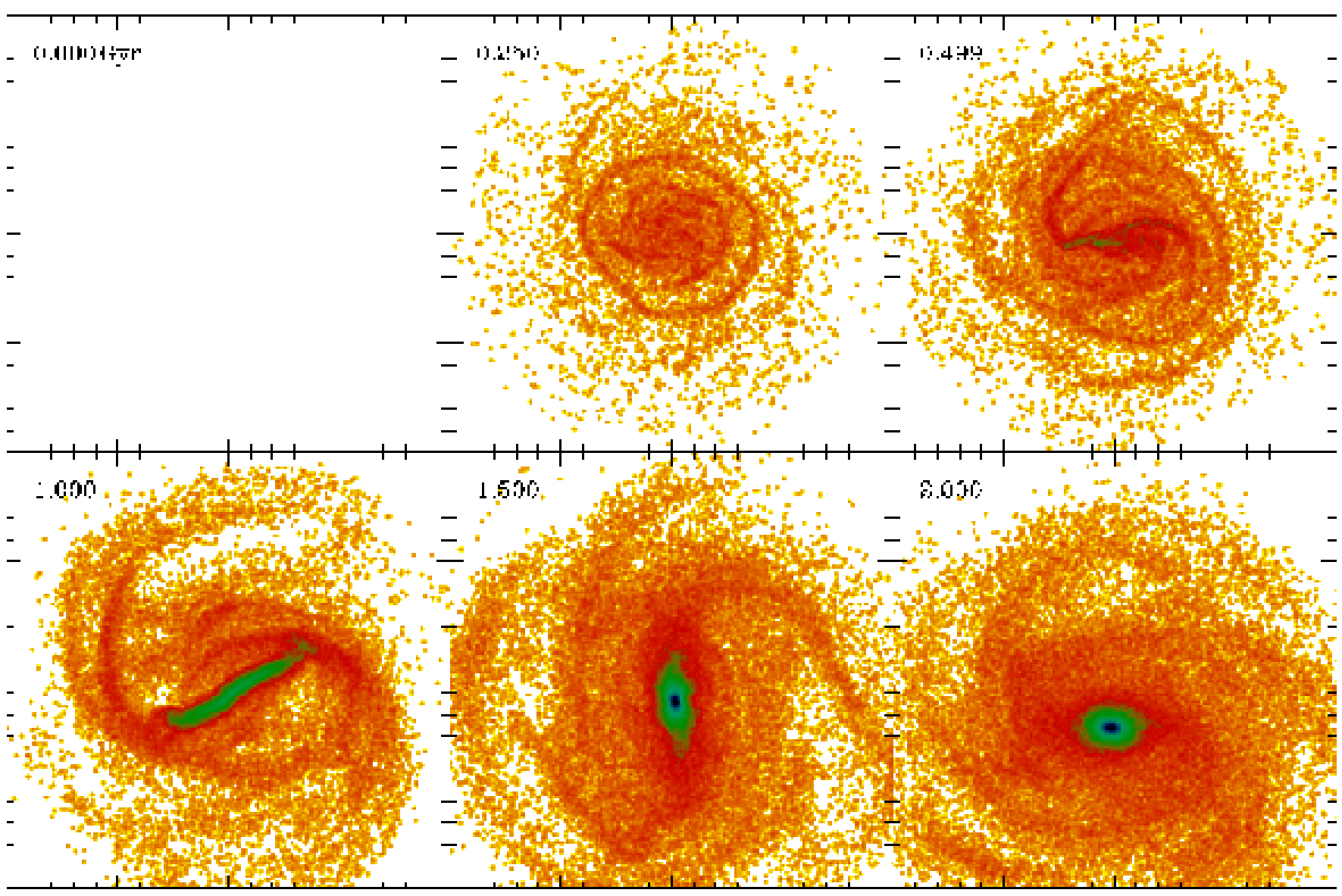}
\caption{Projected density for the new stars formed in the SpGKS8 simulation out to 2 Gyr.  The length of the box is 10 kpc.  The first snapshot is blank as there is no star formation at $t=0$.}
\end{center}
\end{figure}

We also include the SpGKS8 case, with a gas consumption time scale of 8 Gyr.  Figures 18-20 display the projected images stellar, gas, and new stars respectively, which are similar to the SpGKS simulation, but with a slightly greater spiral disturbance seen in Figure 18 at $t=1.5~\rm Gyr$, which is also reflected in the late time $m=2$ Fourier component shown in Figure 21.  The somewhat longer (by $\sim 2$) gas consumption time scale for this simulation allows the gaseous component to couple dynamically to the stellar disk for a slightly longer time.  There is likely to be continued accretion of gas from the intergalactic medium, which may serve to replenish the spiral instability over even longer timescales, a point that we discuss qualitatively in \S 5.  Finally, Figure 24 display the time variation of the pattern speed for the SpGKS8 simulation, which is on average $27~\rm km/s/kpc$ and varies by $\sim 40$\% over 3 Gyr. 

\begin{figure}
\begin{center}
\includegraphics[scale=0.4]{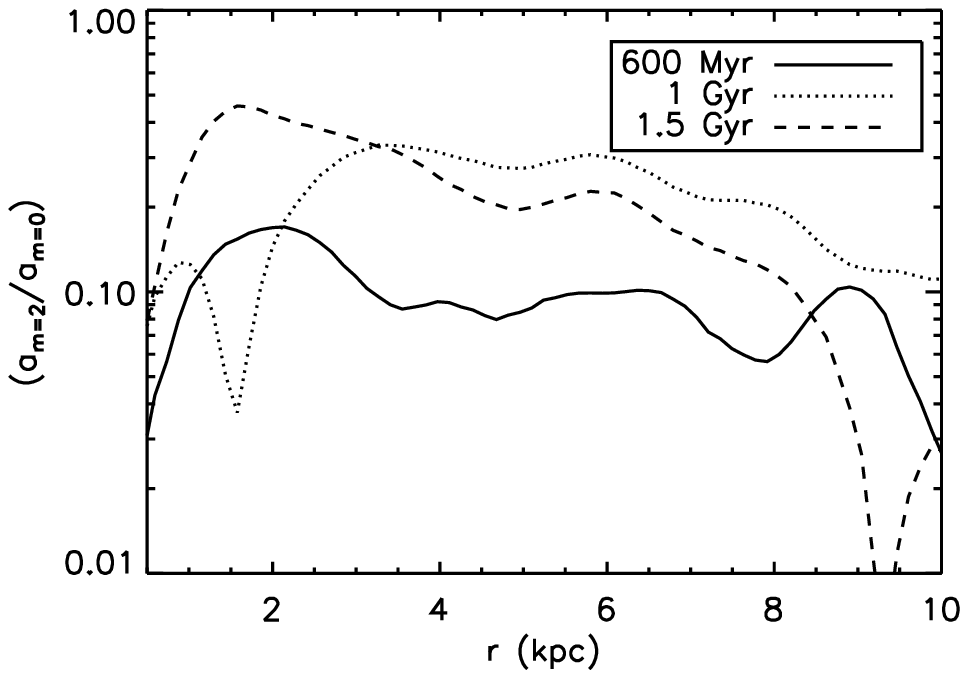}
\caption{The $m=2$ Fourier component of the projected stellar density at three time snaps for the SpGKS8 simulation.}
\end{center}
\end{figure}

\begin{figure}
\begin{center}
   \includegraphics[scale=0.4]{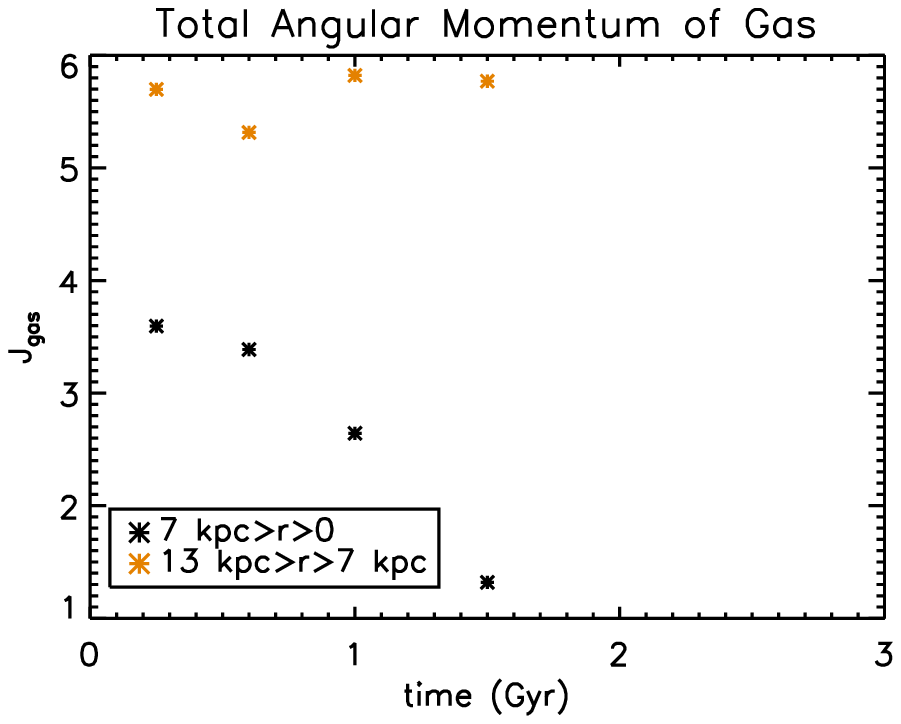}
   \includegraphics[scale=0.4]{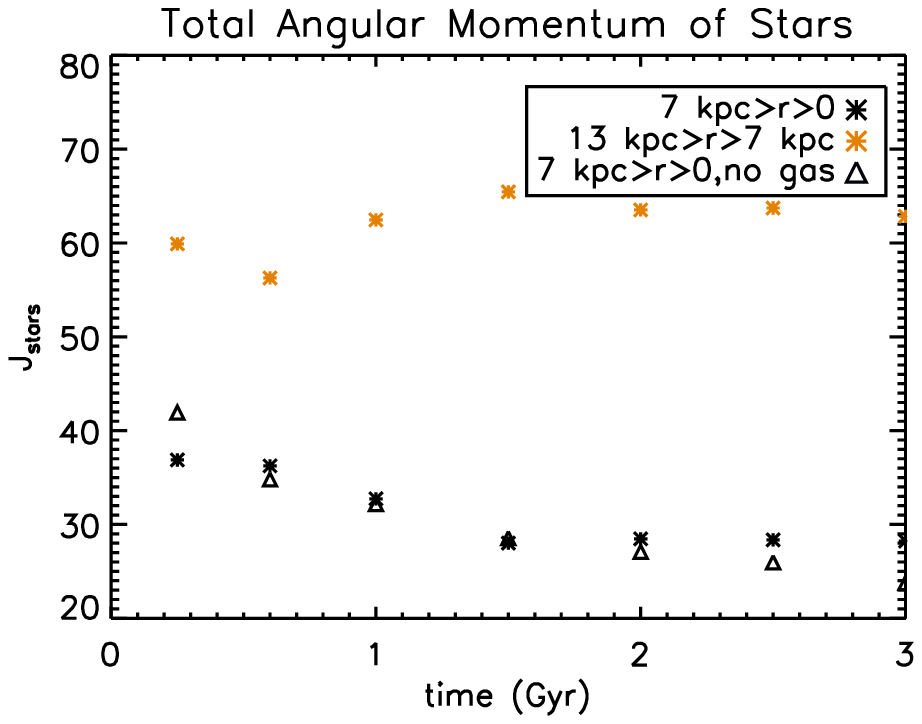}
   \includegraphics[scale=0.4]{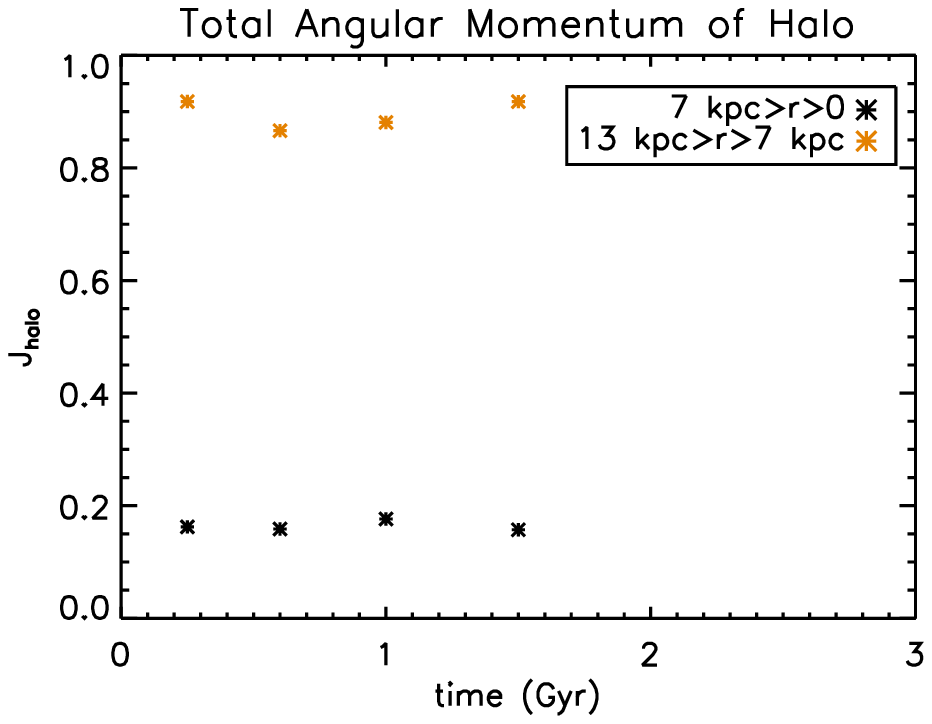} 
   \caption{The time evolution of $J_{z}$ for (a) the gas, (b) the stars overplotted with the reference SpNGNB case (in this case the stars lose 23 \% of their angular momentum inside of $7~\rm kpc$ out to 3 Gyr in contrast to the 43 \% loss seen in the purely collisionless case), and (c) the halo for the SpGKS8 simulation within $0<r<7~\rm kpc$ and $7~\rm kpc<r<13~\rm kpc$}
\end{center}
\end{figure}

\begin{figure}
\begin{center}
\includegraphics[scale=0.4]{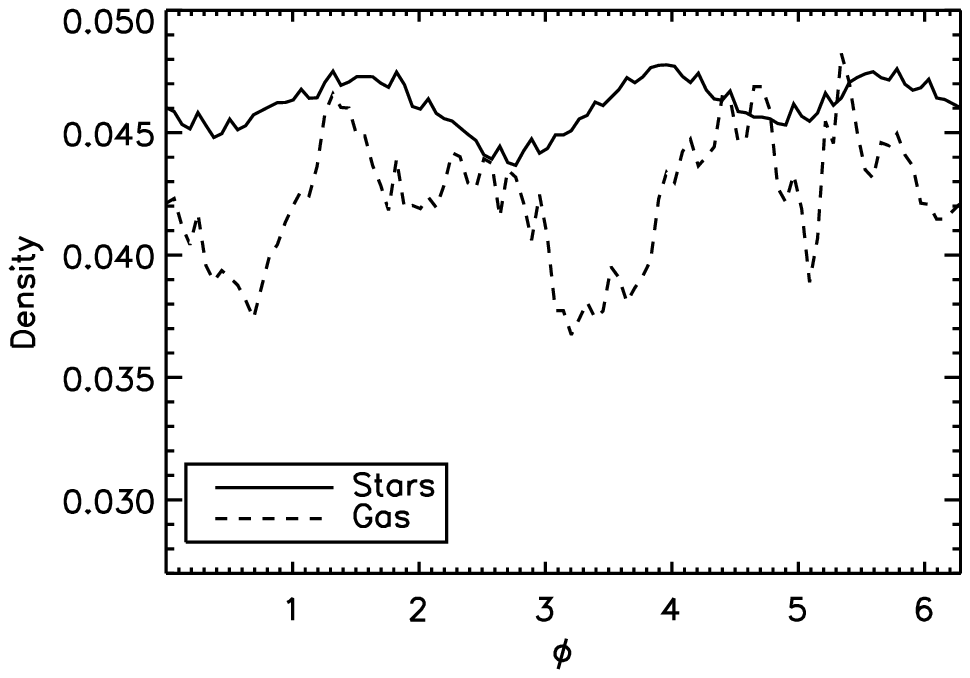}
\caption{Azimuthal phase shift in density (arbitrarily scaled) between the gas and the stars in the SpGKS8 simulation at co-rotation at 250 Myr.}
\end{center}
\end{figure}

\begin{figure}
\begin{center}
\includegraphics[scale=0.4]{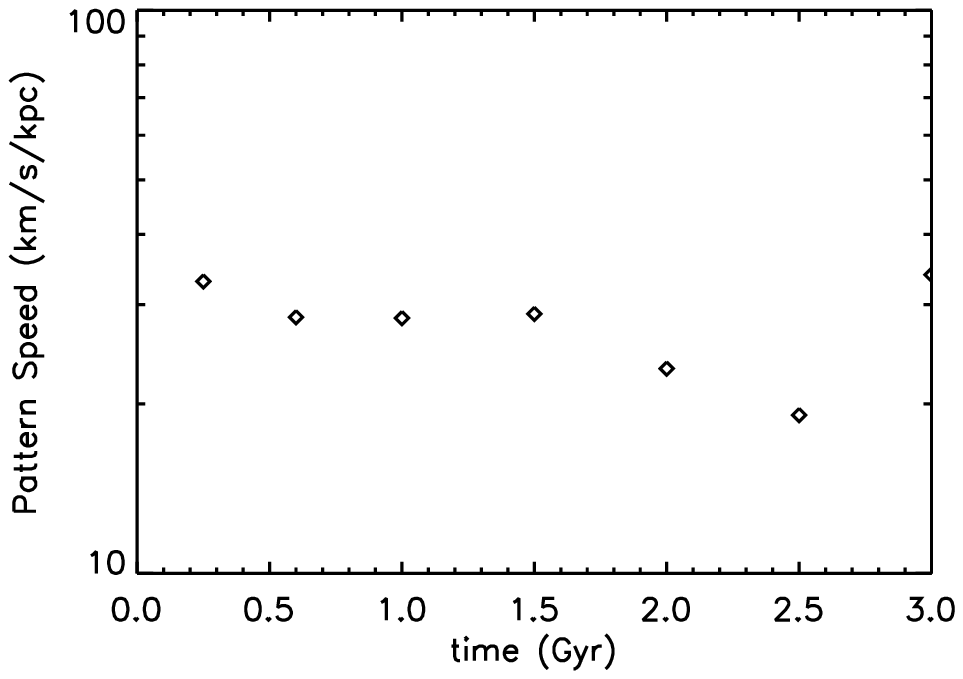}
\caption{The time variation of the pattern speed, out to 3 Gyr (identified using Tremaine \& Weinberg 1984) for the SpGKS8 simulation}
\end{center}
\end{figure}

\section{Discussion}

A number of papers have discussed the role of gas in providing a dissipative and therefore a saturation mechanism (Roberts \& Shu 1972; Lubow, Balbus \& Cowie 1986; Bertin et al. 1998).   Our time-dependent simulations treat the gas-star interplay self-consistently.  The gaseous component is able to cool radiatively and dissipate energy, and thereby remain cool enough to be prone to the spiral instability; the azimuthal phase shift between the gas and the stars allows for the donation of angular momentum from the gas to the stellar disk.  We find that this expropriation of angular momentum is a critical factor in the long-term persistence of spiral structure in the stellar disks of galaxies.  Specifically, the torques that the gaseous spiral sub-structure (those features that are not azimuthally symmetric) exert on the stellar disk transfer angular momentum and therefore lead to a roughly time-steady spiral structure in the stellar disk.  The inclusion of a cold gaseous component also reduces the effective Q parameter of the two-fluid system (Jog \& Solomon 1984; Bertin et al. 1998; CLS03).  Springel et al. (2005) elucidated the effects of the Springel \& Hernquist (2003) multiphase model by performing a parameter space survey; due to the energy injection from supernovae that is implicit in this model, extremely high gas fractions are not unstable to spiral structure.  We discuss this point further in Appendix B.

Kendall et al. (2008) recently produced a composite surface density map of M81 using $\it{Spitzer's}$ IRAC 3.6 $\micron$, 4.5 $\micron$, 8 $\micron$, optical, B, V, I and 2MASS $K_{\rm s}$ band data.  They extract the amplitude and phase of the density from the map, and find that the response of the gas, which is traced by dust emission in the 8 $\micron$ IRAC band, is phase-shifted in azimuth relative to the stellar density wave.  Our analysis of the simulations here is consistent with their findings, with azimuthal offsets between the gas and stars of order one radian.  In a future paper, we perform radiative transfer calculations through the simulations to calculate simulated images as done previously by Chakrabarti \& Whitney (2008) for ULIRGs in the continuum bands.  Our gas surface density images (Figure 6 for example) show a much sharper response as seen in B-band images of spirals, while the stellar surface density images (Figure 5 for example) show a much smoother response as seen in the K band images of local spirals (Block et al. 1996; Elmegreen \& Elmegreen 1984; Rix \& Zaritszky 1995).  

A major caveat to our simulations here is that we have not included the cosmological infall of gas, which can potentially serve to replenish the supply of gas over Gyr timescales as the pre-existing supply of gas is steadily exhausted by star formation and transported inwards by gravitational torques.  Colavitti et al. (2008) have recently analyzed the chemical evolution of a spiral galaxy in a cosmological context and find a gas infall prescription that reproduces observed chemical properties.  Such a prescription may also be used to study the long term dynamics of spiral structure in a more cosmologically motivated way than we have done here.  As we have studied galaxies evolving in isolation which cannot make use of the continued accretion of gas to replenish the critical supply of a cold dissipative component, as gas is tranported inwards or consumed by star formation, we very likely find a lower limit ($\sim 2~\rm Gyr$) for the lifetime of spiral structure in disk galaxies.  This time scale may well be extended by fresh infall of gas.  Gas-rich mergers will certainly affect the stability of disks; however, detailed knowledge of their frequency, mass ratio (i.e., minor or major mergers), and final dynamical impact cannot be obtained without performing high resolution hydrodynamical cosmological simulations.  Recent observational works estimate that a third of LIRGs ($L_{\rm IR} \ga 10^{11} L_{\odot}$) are produced by major mergers at $z \sim 1$, while the rest are morphologically akin to spirals (Elbaz et al. 2007).  The morphological variation of normal galaxies with redshift (due to the influence of other factors such as mergers) is another issue that we have not treated here.  Our approach here has been to study the simple case of a galaxy evolving in isolation and to understand specifically the gas-star dynamical coupling that is a critical factor in the long term persistence of spiral structure in stellar disks.  Due to the relative simplicity of our approach, we have been able to study aspects of the gas physics that impact the angular momentum transfer from the gas to the stars, such as the effect of star formation (by studying simulations where star formation is turned off, i.e., in \S 4, as well as simulations where star formation is turned on and the gas consumption timescale is varied, i.e., in \S 4.1), as well as the effect of the artificial bulk viscosity (Appendix A), and the equation of state (Appendix B).   

\section{Conclusion}

$\bullet$ We find that angular momentum exchange between the self-gravitating gas disk and the stars leads to relatively long-lived spiral structure in the stellar disk, with spiral structure lifetimes of order a few Gyr.  Spiral structure is washed out, when it is incipient, in simulations that do not include a cold, dissipative gaseous component.    

$\bullet$ The gas disk is able torque the stellar disk and donate angular momentum interior to co-rotation.   The gas is able to cool radiatively and dissipate energy - leading to a time lag (or azimuthal phase shift on the order of $\sim 1~\rm radian$ close to co-rotation) between the response of the gas and the response of the stars.  Even though the gas mass is less than the stellar mass, the gas can exert appreciable torques (and donate non-negligible angular momentum) as the azimuthal response which is highly asymmetric (or azimuthal gradient of the potential) is much larger than that of the stars.  Over timescales of many Gyr, there is radial inflow of gas which ultimately limits the donation from the gas (if it is not replenished due to cosmological infall); star formation (depending on gas consumption time-scales) will convert the cold, dissipative component to stars which heat up and have no cooling vent.  

$\bullet$ The primary (specifically the $m=2$) Fourier components of the stellar surface density are roughly time-steady in simulations that include a gaseous component, with variation in surface density of $\sim 3$ over a Gyr.  The pattern speed of the $m=2$ component is more time steady in simulations that include gas than those that do not.

$\bullet$ The halo is a net sink of angular momentum, as has been seen in previous studies of bar instabilities with live halos.

\bigskip
\bigskip
\section*{Acknowledgements}

I am grateful to Frank Shu for insightful discussions on the long-term persistence of spiral structure.  I thank T.J. Cox for advice on performing GADGET simulations and providing a script to plot surface density outputs.  I also thank C.K. Chan, Mark Reid, Phil Chang, and Ramesh Narayan for detailed comments on the manuscript, and Doug Lin, Martin Weinberg, Lars Hernquist, and Dusan Keres for helpful discussions.   

\appendix

\section{Angular Momentum Transfer \& Artificial Bulk Viscosity}

We discuss here the consequence of varying the artificial bulk viscosity parameter and its effect on angular momentum transfer from the gas to the stars.  Figure 23 (a-c) depicts the angular momentum transfer for the SpGNSFR simulation with the bulk viscosity parameter set to zero.  This is similar to the collisionless case, and results in the angular momentum of the stars steadily declining, as there is no transfer from the gas due to a lack of phase shift between the gaseous and stellar components.  The standard value for the artificial bulk viscosity that we have used in the paper is 0.75; this case is shown in Figures 24 (a-c).  Figures 25 (a-c) depict the SpGNSFR case with a artificial bulk viscosity of unity, which results in a very similar behavior of angular momentum transfer as our standard case.  The zero viscosity simulation (SpGNSFR0) shows a 24 \% loss in the stellar angular momentum out to 1.4 Gyr inside of 7 kpc (which is comparable to the 30 \% loss in stellar angular momentum for the reference SpNGNB purely collisionless case), while the standard viscosity case (SpGNSFR) and the SpGNSFR1 both experience a 6 \% loss out to 1.4 Gyr interior to 7 kpc.  In summary, the effect of lowering the artificial bulk viscosity is to approximate the purely collisionless problem, thereby removing the phase shift in azimuth that is needed for the gas to exert a torque on the stellar component.  Aside from the effect of viscosity, gas particles also exert pressure -- this will not affect the angular momentum transport directly, but will affect the propensity of a gas disk to be prone to spiral structure.  We discuss the effect of pressure (i.e., assumptions about the equation of state) in Appendix B.

\begin{figure}
\begin{center}
   \includegraphics[scale=0.4]{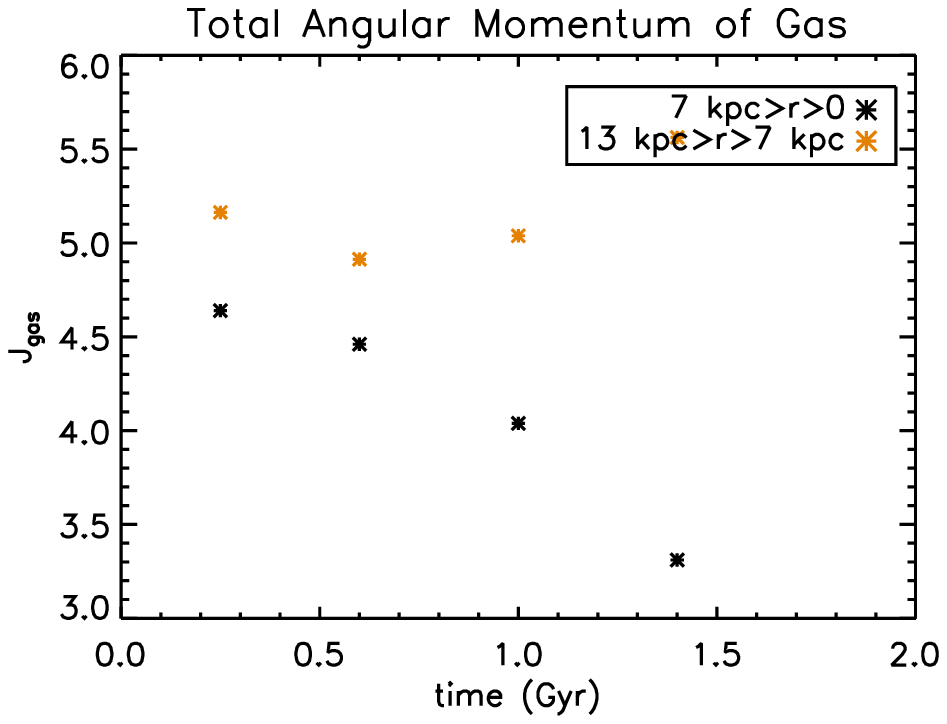}
   \includegraphics[scale=0.4]{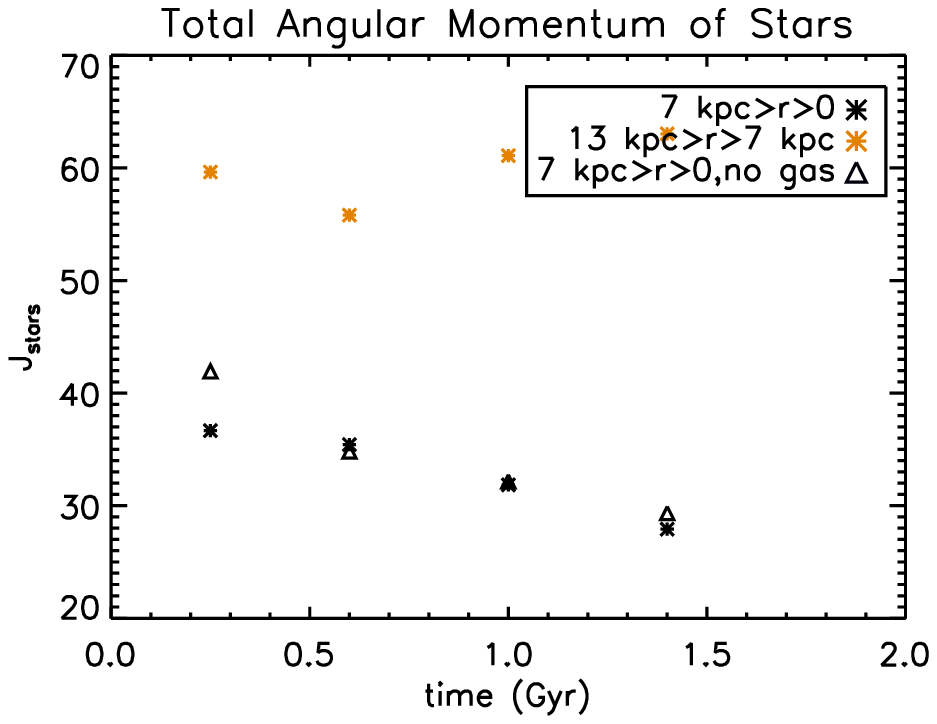}
   \includegraphics[scale=0.4]{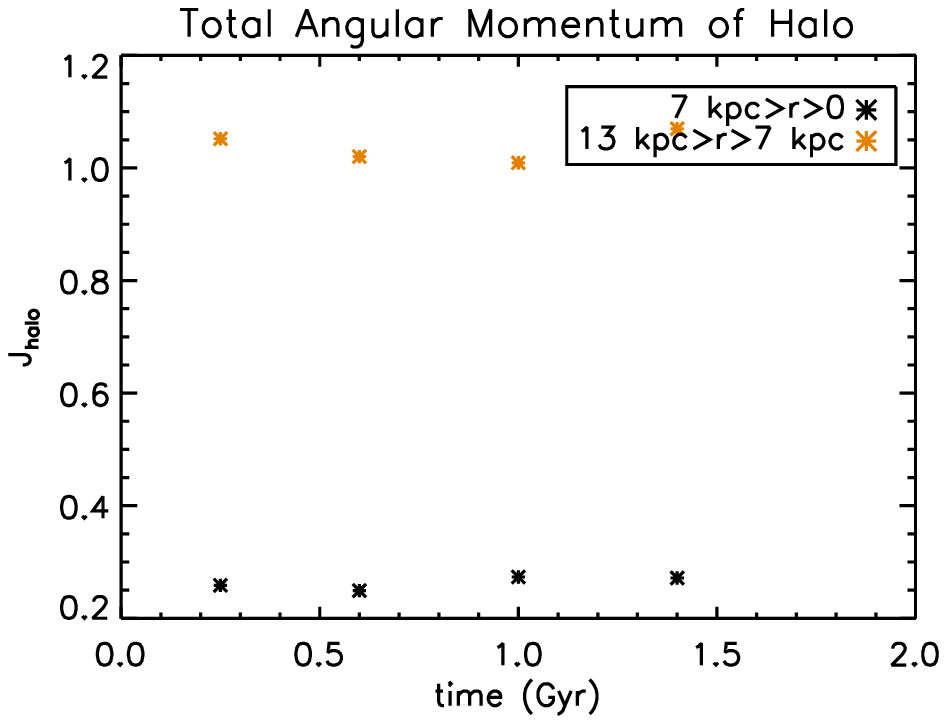} 
   \caption{J transfer for SpGNSFR simulation with the bulk artificial viscosity set to zero (a) gas (b) stars, overplotted with the reference SpNGNB case which has a 30 \% loss in angular momentum out to 1.4 Gyr interior to 7 kpc compared to the 24 \% loss for this zero viscosity case, and (c) halo}
\end{center}
\end{figure}

\begin{figure}
\begin{center}
   \includegraphics[scale=0.4]{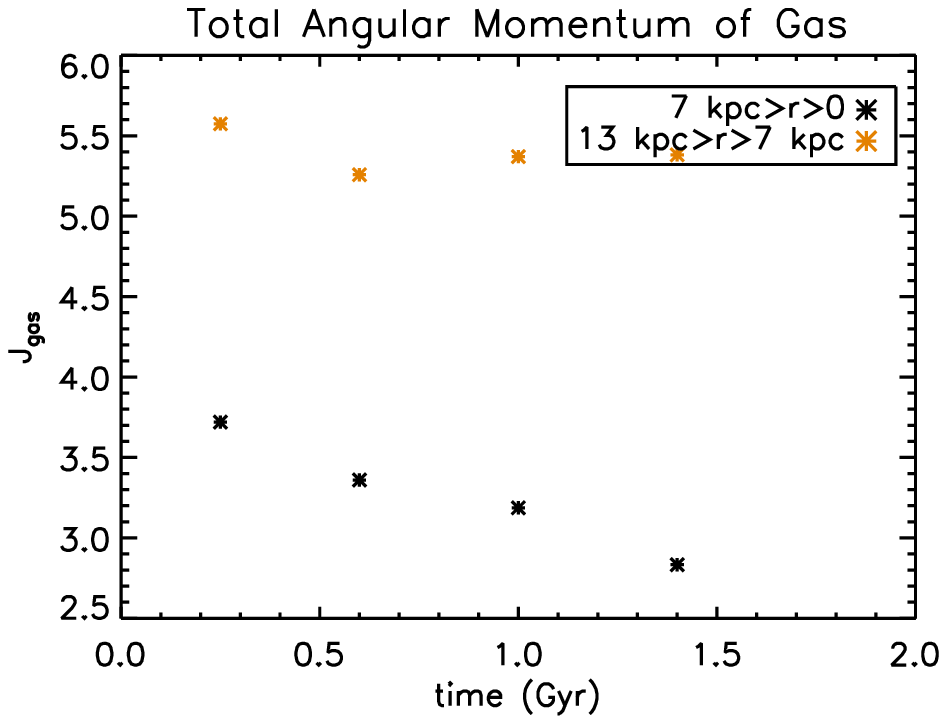}
   \includegraphics[scale=0.4]{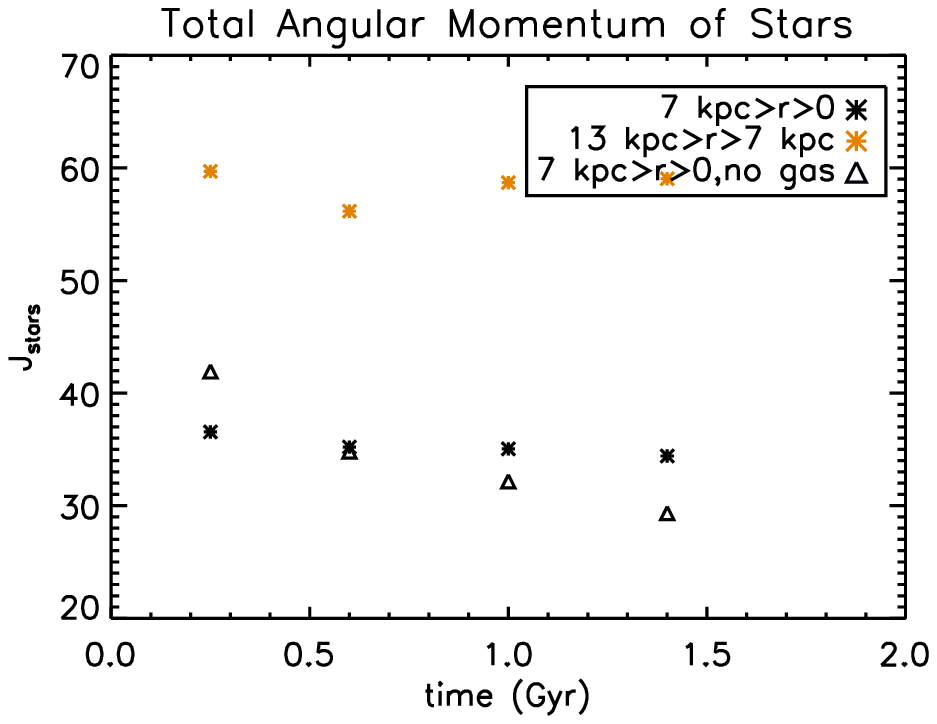}
   \includegraphics[scale=0.4]{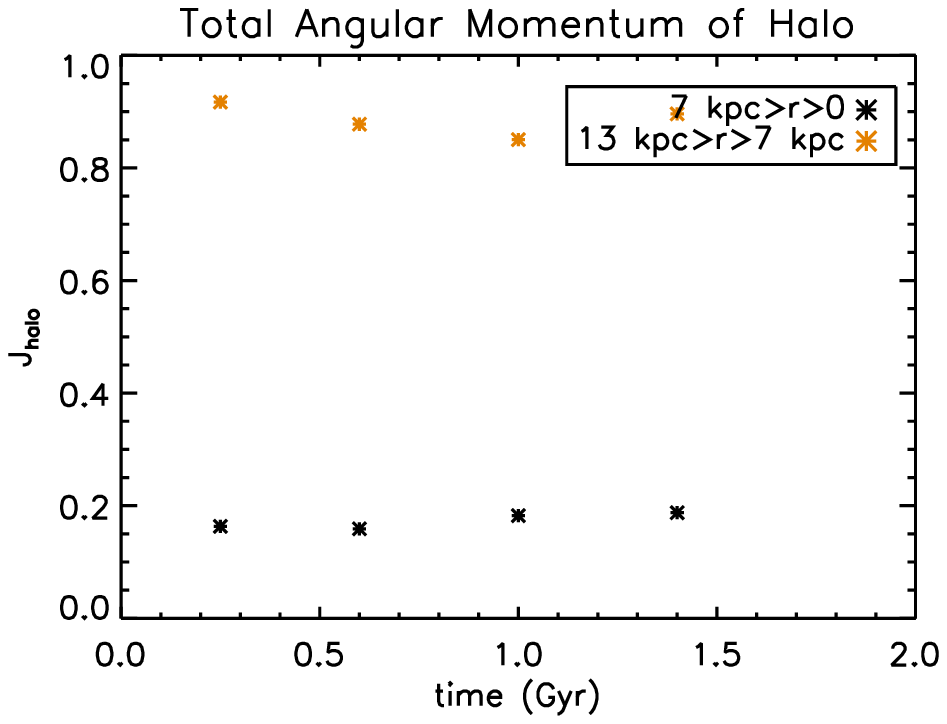} 
   \caption{The time evolution of $J_{z}$ for (a) the gas, (b) the stars overplotted with the reference SpNGNB case which has a 30 \% loss in angular momentum in contrast to the 6 \% loss seen in this standard viscosity case, and (c) the halo for the SpGNSFR simulation within $0<r<7~\rm kpc$ and $7~\rm kpc<r<13~\rm kpc$.  This is the standard case with the bulk artificial viscosity set to 0.75.}
\end{center}
\end{figure}

\begin{figure}
\begin{center}
   \includegraphics[scale=0.4]{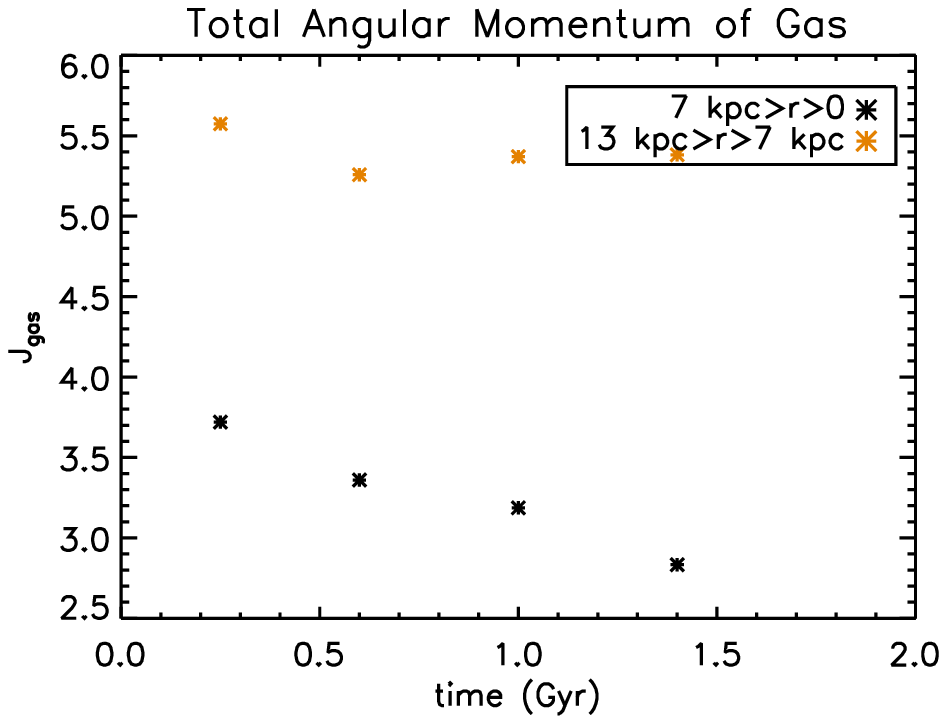}
   \includegraphics[scale=0.4]{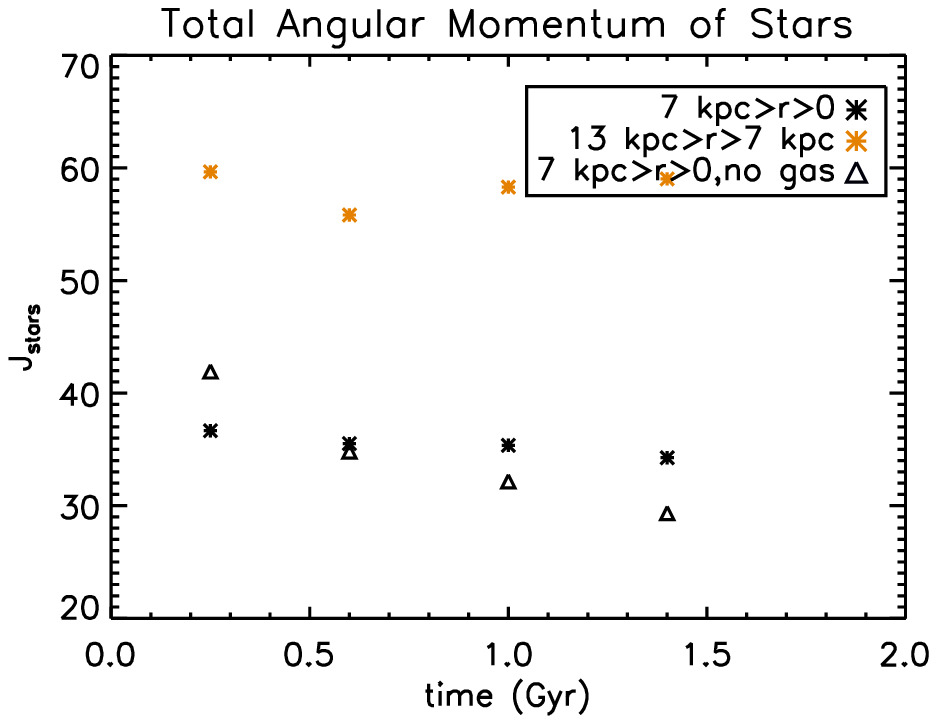}
   \includegraphics[scale=0.4]{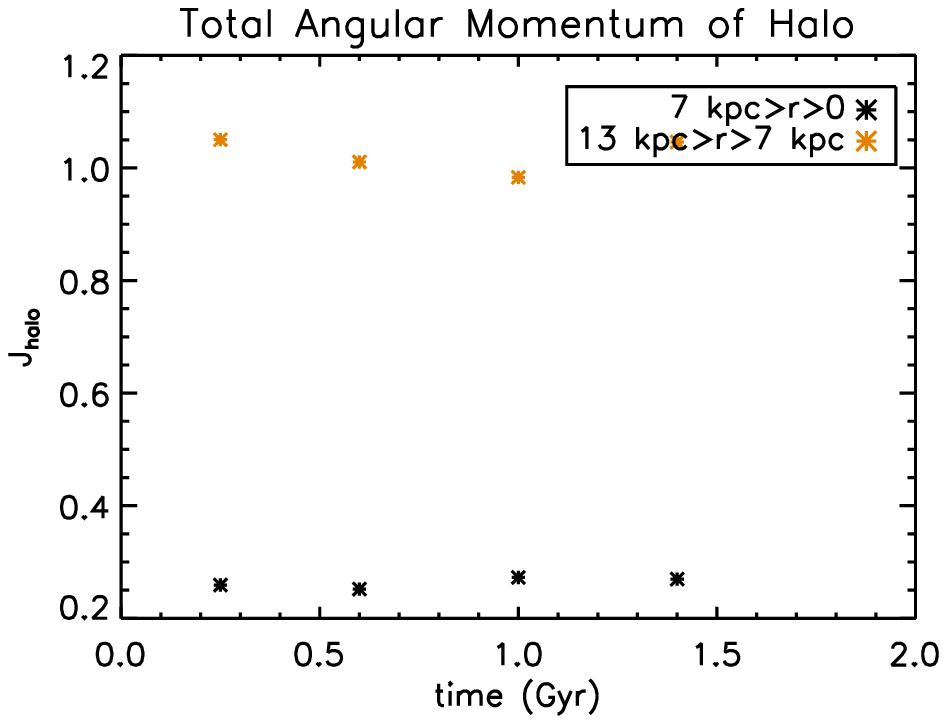} 
   \caption{J transfer for SpGNSFR simulation with the bulk artificial viscosity set to one (a) gas (b) stars overplotted with the reference SpNGNB case which has a 30 \% loss out to 1.4 Gyr in contrast to the 6 \% loss in angular momentum seen here, and (c) halo}
\end{center}
\end{figure}

\clearpage

\section{Equation of State}

Another aspect of the gas physics that affects the stability of the gas disk, and indirectly impacts the angular momentum transport to the stellar disk is the assumed equation of state for the gas.  As discussed by Springel et al. (2005), the equation of state, when it is taken to be greater than zero which is the isothermal limit, models the effect of supernova feedback by over-pressurizing regions relative to an isothermal equation of state.  In the isothermal limit, the pressure in compressed regions increases in proportion to the density.  The multiphase model mediates that for $\rho > \rho_{\rm th}$, the gas is thermally unstable and stars can form; the energy injection rate from supernovae in this model is proportional to the star formation rate.  The compressed gas then would have an effective equation of state that is stiffer than isothermal if it retains this feedback energy.  One can expect that simulations performed with equations of state that are stiffer than isothermal are somewhat less prone to a spiral instability as the effective sound speed has been increased.  We vary the equation of state from 0-1 in Figures 26(a-c) and find that the simulation performed with the isothermal equation of state (EQS0LR) shows the lowest decrease in stellar angular momentum, while the full multiphase model (EQS1LR) has a 25 \% decrease in the stellar angular momentum out to 1.4 Gyr.  However, this relative decease -- 18 \% versus 25 \%, as the equation of state varies from isothermal to the fully pressurized multiphase model is a small enough change to suggest that assumptions about the equation of state for the gas only slightly affect the angular momentum transport to the stellar disk, for a gas fraction of 10 \%.  Springel et al. (2005) have performed a comprehensive parameter study in terms of the gas fraction and the equation of state to study the effects these parameters have on the gaseous disk.  They find that for the full multiphase model (stiff equation of state) even pure gas disks are stable to fragmentation, with an increase in gas fraction lowering the susceptibility of the gas disk to spiral structure.  In the multiphase model, increasing the gas fraction increases the star formation (as mediated by the adopted Kennicutt-Schmidt algorithm) and thereby the supernova feedback energy (which is proportional to the star formation rate) that is deposited into the gas.  This can cause the effective temperature of the gas to be dynamically hotter than the stars.  On the other exterme, extremely high gas fraction disks that are evolved with an isothermal equation of state undergo violent fragmentation.  Thus, extremely high gas fraction disks (whether they are evolved with an isothermal equation of state or full multiphase model) will not be effective as angular momentum donors to the stellar disk.

\begin{figure}
\begin{center}
   \includegraphics[scale=0.4]{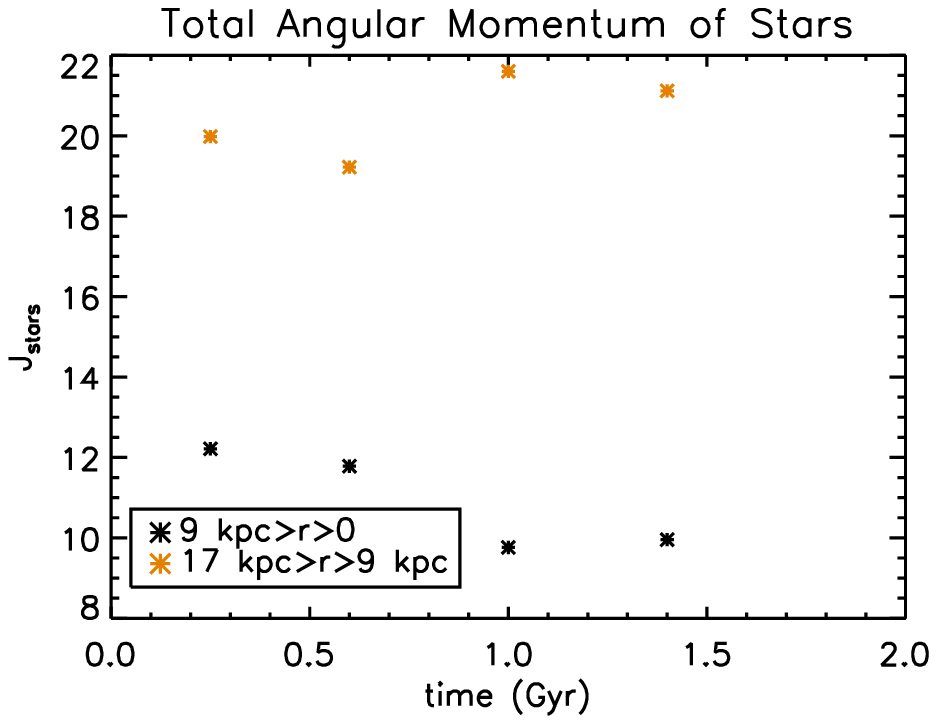}
   \includegraphics[scale=0.4]{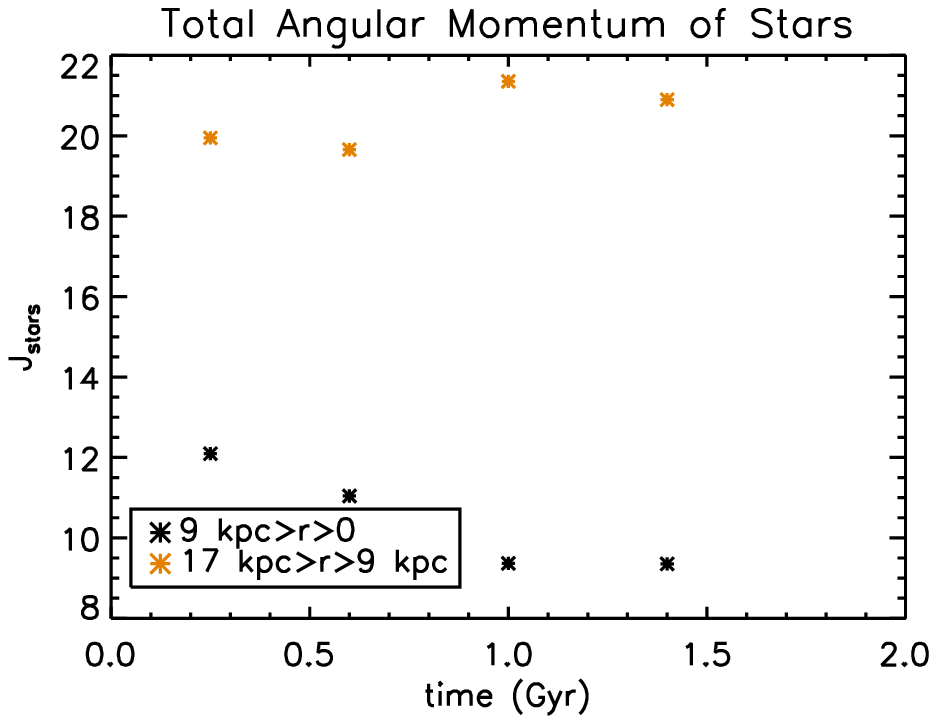}
   \includegraphics[scale=0.4]{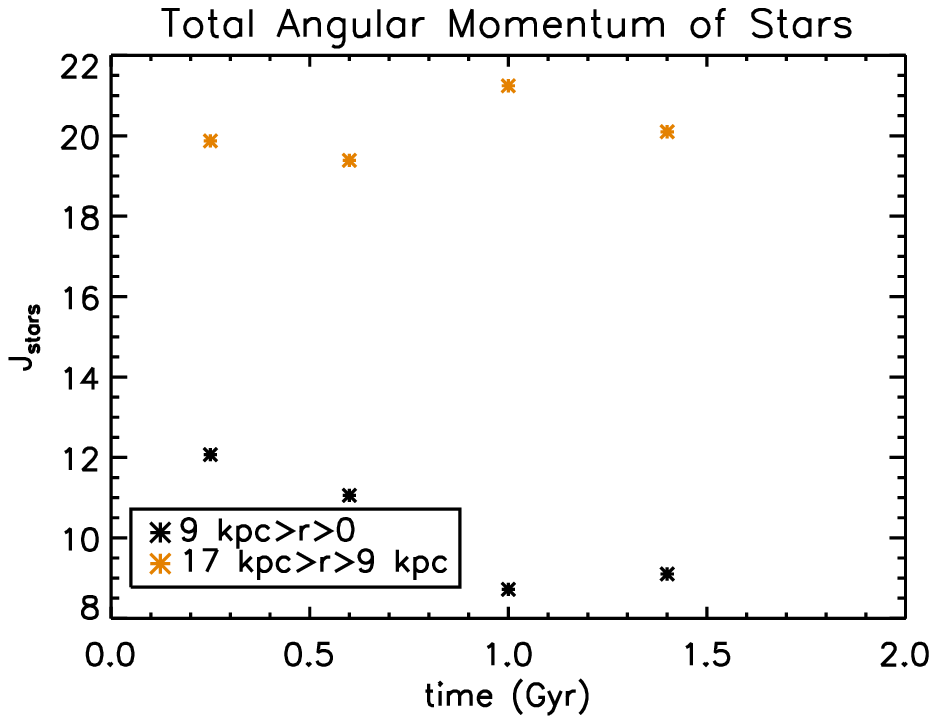} 
   \caption{Stellar angular momentum as a function of time for (a) the EQS0LR simulation with the equation of state to zero (isothermal), which shows a 18 \% loss out to 1.4 Gyr (b) for the EQSSTLR simulation with the equation of state set to 0.25 (our standard case which is over-pressurized relative to the isothermal case); this has a 22 \% loss in the stellar angular momentum out to 1.4 Gyr (c) for the EQS1LR simulation with the equation of state set to one that is the full multi-phase model shows a 25 \% loss in angular momentum out to 1.4 Gyr}
\end{center}
\end{figure}

\clearpage
% ####### Bibliography ########################################################

% ######## End bibliography ###################################################
\label{lastpage}


\begin{thebibliography}{99}
\bibitem[]{}Athanassoula,E., 2003, MNRAS, 341, 1179A  
\bibitem[]{}Bertin,G. \& Lin, C.C., 1999, Spiral Structure in Galaxies: A Density Wave Theory, Cambridge MA MIT Press, ISBN0262023962  
\bibitem[]{}Balsara,D.S., 1995, Journal of Computational Physics, Vol. 121, no. 2, p. 357-372 
\bibitem[]{}Berentzen,I., et al., 2007, ApJ, 666, 189B 
\bibitem[]{}Block,D.L.,Bertin,G.,et al., 1994, A\&A,288,365B  
\bibitem[]{}Chakrabarti,S., Laughlin,G., \& Shu, F.H., 2003, ApJ, 596, 220C (CLS)  
\bibitem[]{}Colavitti,E.,et al., 2008, A\&A,483,401C 
\bibitem[]{}Elmegreen,D.M. \& Elmegreen,B.G.,1984,ApJS,54,127E 
\bibitem[]{}Lin, C.C. \& Shu, F.H., 1964, ApJ, 140, 646L  
\bibitem[]{}Lin,C.C., Yuan, C. \& Shu, F.H., 1969, ApJ, 155, 721 L 
\bibitem[]{}Goldreich,P. \& Lynden-Bell, D., 1965, MNRAS, 130, 125G  
\bibitem[]{}Hernquist, L. \& Mihos, C., 1995, ApJ, 448, 41 H  
\bibitem[]{}Hohl,F., 1970, NASTR.343,H  
\bibitem[]{}Kendall,S., et al., 2008, MNRAS, 387, 1007K 
\bibitem[]{}Kennicutt,R.C. 1983, ApJ, 272, 54 
\bibitem[]{}Kim,W.T. \& Ostriker, E.C.,2002,ApJ,570,132K  
\bibitem[]{}Kim,W.T. \& Ostriker, E.C.,2007,ApJ,660,1232K  
\bibitem[]{}Laughlin,G.,Korchagin,V. \& Adams,F.C.,1998,ApJ,504,945L  
\bibitem[]{}Lubow,S.H. et al., 1983, ApJ, 309, 496L 
\bibitem[]{}Lynden-Bell,D. \& Kalnajs, A.J., 1972, MNRAS,157, 1L 
\bibitem[]{}Martinez-Valpuesta,I., et al., 2006, ApJ, 637, 214 M 
\bibitem[]{}Rix,H. \& Zaritsky,D., 1995, ApJ, 447, 82R 
\bibitem[]{}Roberts, W.W. \& Shu, F.H., 1982, ApL, 12, 49 R 
\bibitem[]{}Sellwood,J. \& Carlberg, R.G., 1984, ApJ, 282, 61S  
\bibitem[]{}Shetty,R. \& Ostriker, E.C., 2006, ApJ, 647, 997S (S006)   
\bibitem[]{}Shu, F.H., et al., 1971, ApJ, 166, 465S 
\bibitem[]{}Shu, F.H., Milione, V. \& Roberts, W., 1973, ApJ, 183, 819S (SMR) 
\bibitem[]{}Shu, F.H., et al., 2007, ApJ, 662L, 75S 
\bibitem[]{}Springel,V., et al., 2001, New Astronomy, Vol. 6, Issue 2, p. 79-117. 
\bibitem[]{}Springel,V., 2005, MNRAS, 364, 1105S 
\bibitem[]{}Thilker, D.A., et al., 2007, ApJS, 173, 538T 
\bibitem[]{}Toomre, A., 1964, ApJ, 139, 1217T  
\bibitem[]{}Toomre, A., 1981, The structure and evolution of normal galaxies; Proceedings of the Advanced Study Institute, Cambridge, England, August 3-15, 1980. (A82-11951 02-90) Cambridge and New York, Cambridge University Press, 1981, p. 111-136  
Tremaine,S. \& Weinberg, M.D., 1984, ApJ, 282L, 5T 
\bibitem[]{}Wong,T. \& Blitz,L., 2002, ApJ, 569, 157W 
\bibitem[]{}Wong,T., Blitz,L. \& Bosma, A., 2004, ApJ, 605, 183 W 
\bibitem[]{}Yanez,M., Norman,M., et al., 2007, arXiv:0710.1331  

\end{thebibliography}
\end{document}